\documentclass[12pt]{article}

\usepackage{amsfonts,ulem}
\usepackage{amsmath}
\usepackage{amssymb,authblk}
\usepackage{amsthm}
\usepackage{array}
\usepackage{graphicx}
\usepackage[margin=1in]{geometry}
\usepackage{float}
\usepackage{booktabs}
 
\usepackage[config,labelfont=sc,textfont=sl]{caption,subfig}

\usepackage{mathtools}
\usepackage{psfrag}
\usepackage{mathrsfs}
 
\newcommand{\rmd}{\mathrm{d}} 
\newcommand{\rme}{\mathrm{e}}

\newcommand{\rmi}{\mathrm{i}}

\newcommand{\rmr}{\mathrm{r}}

\newcommand{\rmB}{\mathrm{B}} 
 
\newcommand{\rmD}{\mathrm{D}}

\newcommand{\rmH}{\mathrm{H}} 
 
\newcommand{\rmJ}{\mathrm{J}} 
 
\newcommand{\rmL}{\mathrm{L}}

\newcommand{\rmT}{\mathrm{T}}  
\newcommand{\rmU}{\mathrm{U}}

\newcommand{\rmY}{\mathrm{Y}}

  \newcommand{\bfk}{\mathbf{k}}  
    \newcommand{\bfA}{\mathbf{A}} 
        \newcommand{\bfB}{\mathbf{B}}   
      \newcommand{\bfg}{\mathbf{g}}  
      \newcommand{\bfR}{\mathbf{R}}   
       \newcommand{\bfK}{\mathbf{K}}   
         \newcommand{\bfQ}{\mathbf{Q}}

\newcommand{\I}{\mathrm{i}} 
\newcommand{\E}{\mathrm{e}}

\title{\vspace{-20mm} Two-dimensional Helmholtz resonator arrays. Part I. Matched asymptotic expansions  for thick- and thin-walled resonators}

\author[1]{M. J. A. Smith}
\author[1]{I. D. Abrahams} 

\affil[1]{\small Department of Applied Mathematics and Theoretical Physics, University of Cambridge,
Wilberforce Road, CB3 0WA, UK}

\date{}





\begin{document}
\maketitle
  \vspace{-10mm}
\begin{center}
\small {\it Submitted Manuscript}
\end{center}	
\begin{abstract}
We present a novel multipole formulation for computing the band structures of two-dimensional arrays of cylindrical Helmholtz resonators. This formulation  is derived by combining  existing multipole methods for arrays of ideal cylinders with   the method of matched asymptotic expansions.  We  construct asymptotically close representations for the dispersion equations of the first band surface, correcting and extending an established lowest-order (isotropic) result in the literature for thin-walled resonator arrays. The descriptions we obtain for the first band are accurate over a relatively broad frequency and Bloch vector range  and not simply in the long-wavelength and low-frequency regime, as is the case in many classical   treatments. Crucially, we are   able to  capture features of the first band, such as low-frequency anisotropy, over a broad range of filling fractions, wall thicknesses, and   aperture angles. In addition to describing the first band we use our  formulation   to compute   the first band gap  for both  thick- and thin-walled resonators, and find that thicker resonator walls correspond to both a narrowing of the first band   gap and an increase in the central band gap frequency.
\end{abstract}

\section{Introduction} 
In recent years,  researchers within the metamaterials and composite materials communities have uncovered a vast array of   media exhibiting interesting and unexpected  wave scattering properties. These have ranged from ultralow frequency band gaps to one-way edge states and  negative refraction \cite{milton2002theory,cui2010metamaterials,xin2020topological},   in a diverse range of wave settings, for example, from acoustics and elasticity through to electromagnetism. The  ongoing development of novel materials remains a very topical and important endeavour for mathematicians, physicists, engineers, and materials scientists alike.    In order to describe compactly  the performance of meta/composite materials, significant attention has been directed towards the efficient calculation of {\it band diagrams} and on obtaining {\it effective medium descriptions}, i.e.,    homogenising the   medium, in a   range of settings. 

From across the literature, a  diverse selection of homogenisation tools have likewise emerged, ranging from fully numerical procedures to   analytical methods that yield  elegant closed-form expressions \cite{milton2002theory,movchan2002asymptotic,cui2010metamaterials}.  One established and well-known analytical procedure  combines multipole methods    with conventional asymptotic methods to obtain closed-form   descriptions  for two-dimensional arrays of cylinders  embedded in a background material  \cite{rayleigh1892lvi,movchan2002asymptotic}. These descriptions for non-resonant arrays of scatterers have proven exceptionally useful for developing highly tuned materials whose properties   lie   between those of the inclusion and matrix phase (analogously to the way an array of resistors combined in series or in parallel form   effective resistances). However, as with the vast majority of effective medium descriptions,   the analytical representations describe the first band surface only   at both low-frequencies and at long-wavelengths. In place of this limited descriptions, it is much more advantageous to obtain descriptions of the first band over a broader range.

In this work, we attempt to obtain  simple asymptotic descriptions of the first band surface over the entire Brillouin zone for a two-dimensional Helmholtz resonator array, and more generally, present a multipole formulation for computing band diagrams over a   wide  frequency range. We use a combination of multipole methods \cite{movchan2002asymptotic,parnell2006dynamic} and  the method of matched asymptotic expansions \cite{crighton1992modern,bender2013advanced,cotterill2015time} to obtain results for an array of thin-walled resonators, deriving and providing a small correction to the  result published  in Llewellyn--Smith \cite{llewellyn2010split},  as well as extending treatments to obtain crucial next-order corrections that capture the anisotropy of the medium. We also derive an analogous formulation for thick-walled resonator arrays and present corresponding results. 

The methods outlined here yield a more   general homogenisation result to those obtained for arrays of cylinders, which describe  the first band surface only at low-frequencies and at long-wavelengths (a relatively small segment of the total first band surface). The   expressions obtained here   for the first band  will prove useful for practical applications, admitting     closed-form expressions for both the  phase and   group velocity  inside the crystal, for example. In addition to capturing Bloch vector and frequency dependence (spatial and frequency dispersion), our descriptions also give   the width of the first (subharmonic) band gap in a   range of resonator array configurations. To the best of our knowledge, we are unaware of such   analytical results for two-dimensional resonant arrays,  although there are close similarities to a lowest-order   result  for {\it thin walled} Helmholtz resonator arrays   \cite{llewellyn2010split}. That said, relatively few  analytical studies of this nature exist  due to the   complexity involved in their derivation, although there is an extensive literature on numerical results (see for example, \cite{hu2005two,li2013tuning} for finite-difference time-domain and finite-element method treatments).  In this work we do not rely upon {\it lumped-element models} or {\it  lumped acoustic elements}, which have been used extensively in the literature to model Helmholtz resonators; such treatments  replace the resonator with an equivalent mass-and-spring  or      circuit, which has proven useful in the past for describing   resonators in the deeply  long-wavelength regime \cite{pierce2019acoustics}.  

The descriptions we obtain for the first band complements other work in the literature on two-dimensional arrays of resonators governed by Helmholtz's equation, such as work on  thick cylindrical resonators possessing multiple apertures    \cite{guenneau2007acoustic,antonakakis2013asymptotics} which exhibit effects such as negative refraction. Other Helmholtz equation  studies of this type include work  on two-dimensional arrays of thick-walled split-ring resonators   \cite{movchan2004split} and two-dimensional arrays of closely-packed solid cylinders   \cite{vanel2017asymptotic}.  Estimates for the upper- and lower-bounds of the first band gap in elastic  resonator array problems  have also been considered \cite{krynkin2013analytical}. Research on resonator arrays has also been conducted extensively for Maxwell's equations, including a numerical studies on determining effective optical constants of two-dimensional array of infinitesimally thin split-ring resonators   \cite{juarez2018magnetic}. Another related area examines arrays of gas bubbles in liquids; the fundamental frequency at which the bubble wall oscillates is analogous to a Helmholtz resonance and induces low-frequency band gaps within the fluid medium    \cite{ammari2017subwavelength}.   There has also been   interest within the water waves community on arrays of graded thin-walled Helmholtz resonators, which  can exhibit  strong field amplification, a feature which may prove   useful in   energy harvesting systems \cite{bennetts2019low}.

In addition to Bloch problems, considerable interest has been focused on two-dimensional   {\it scattering} by  Helmholtz    resonator arrays, including acoustic wave scattering by  thick and thin-walled resonators with multiple apertures and by  split-ring resonators   \cite{moran2016embedding}, acoustic wave scattering by elastic (non-rigid) cylindrical resonator arrays in two dimensions \cite{krynkin2010predictions}, and scattering by finite arrays of thin Helmholtz resonators \cite{montiel2017analytical}. Recently, work in resonant arrays embedded in thin films and interfaces (metasurfaces)  has emerged as an  area of interest \cite{schwan2017sound}, such as that seen with two-dimensional arrays of finite-depth resonators implanted beneath the surface of  a half-space   \cite{brandao2020asymptotic}, as well as one-dimensional arrays of resonators \cite{maling2017radiation,maurel2019enhanced}. Finally,  work   on   arrays of harbours or coves in deep water  are of relevance \cite{monkewitz1985response}, as well as investigations on one-dimensional arrays of    resonators     in thin elastic plates \cite{meylan2017perforated}.
 
 The outline of this paper is as follows. First we present the boundary value problem   for a two-dimensional doubly-periodic array of thin-walled Helmholtz resonators   in Section \ref{chap:probform}. We then set up the matching scheme by examining the field close to an aperture in Section \ref{chap:innerthin}, and derive field asymptotics as we move out from this inner region.  Next, we construct an {\it outer solution} outline in Section \ref{chap:outthin}, where the presence of the small aperture is modelled by a simple source term. We then  conduct asymptotic matching in Section \ref{chap:outthin}\ref{eq:matchedasy}   to obtain our eigensystem  in Section \ref{chap:outthin}\ref{chap:latticeray}. This allows us to derive   the leading-order dispersion equation for the first spectral band in Section \ref{chap:outthin}\ref{eq:leadingordapprox} followed by its first-order correction in Section \ref{chap:outthin}\ref{chap:correction}. In Section \ref{chap:numericals} we consider numerics for a selection of geometries   to demonstrate the efficacy of our approximations. This is followed by a treatment for thick-walled resonators in Section \ref{chap:thicksec}, where we outline all modifications and present additional numerical results. Finally we offer some concluding remarks in Section \ref{chap:concl}.

\section{Problem formulation}
\label{chap:probform}
\begin{figure}[t]
	\centering
	 \subfloat[Subfigure 1 list of figures text][]{
\includegraphics[width=0.41\textwidth]{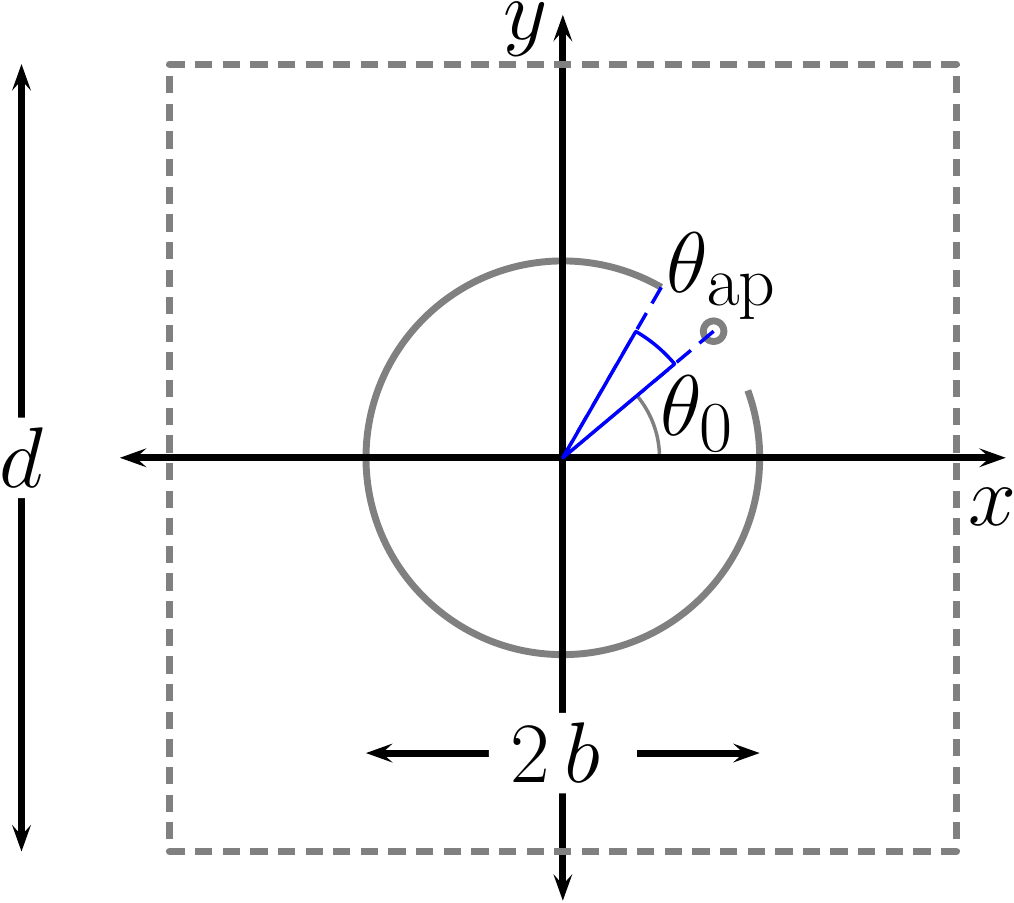}
\label{fig:schemsubfig1}}\hspace{10mm} 
\subfloat[Subfigure 2 list of figures text][]{
\includegraphics[width=0.45\textwidth]{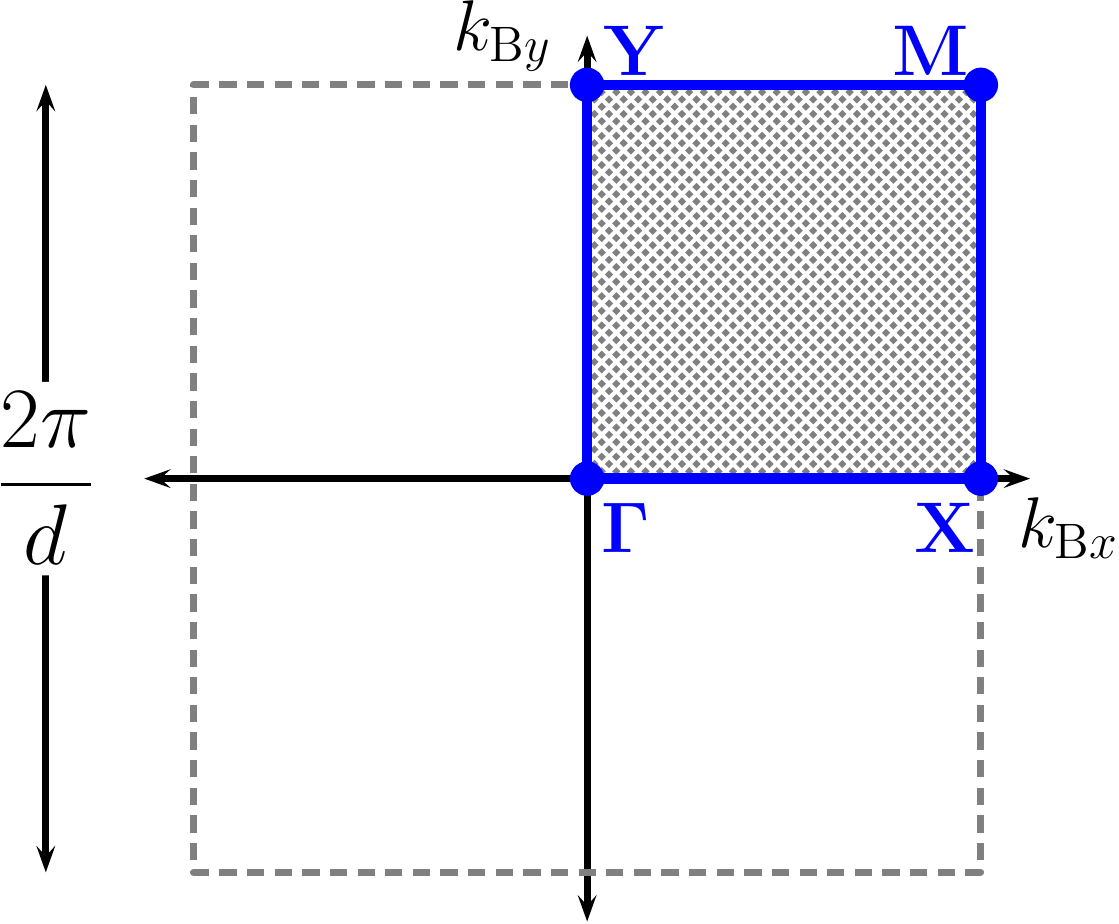}
\label{fig:schemsubfig2}}
	
	\caption{(a) Fundamental (dimensionless) unit cell for a square array of period $d$ containing a thin cylindrical resonator of radius $b$ with   aperture of   arc length   $2\ell$ centred at $(b,\theta_0)$ in polar coordinates, i.e., with half-angle subtended by the aperture  given by $\theta_\mathrm{ap} = \ell/b$; (b) Corresponding unit cell in reciprocal space, for high symmetry values of $\theta_0$, expressed in terms of non-dimensional Bloch coordinates $(k_{\rmB x},k_{\rmB y})$ with irreducible Brillouin zone  shaded, bounded by blue lines, and marked with vertices  $\Gamma = (0,0)$,  $X = (\pi/d,0)$, $Y = (0,\pi/d)$, and $M=(\pi/d,\pi/d)$. \label{fig:resonatorarray}}
\end{figure}

We consider a two-dimensional square array of thin-walled   resonators  spaced a distance $\overline{d}$ apart, that are modelled as  cylinders  of radius $\overline{b}$, each containing an   aperture of   arc length $2\overline{\ell}$   centred about the central angle $\theta_0$. These  are immersed in an acoustic medium of infinite extent  satisfying   the two-dimensional scalar Helmholtz equation    
\begin{equation}
\label{eq:helmnondim}
(\partial_{\overline{x}}^2 + \partial_{\overline{y}}^2)   \, \overline{\phi} +k^2  \overline{\phi} = 0,
\end{equation}
with  Neumann boundary conditions imposed on all  resonator walls. The overbar is used to denote dimensional  quantities, and we take $\overline{\phi}$ to be the steady-state monochromatic field oscillating at angular frequency $\omega$, i.e., the observed time-dependent field is $\mathrm{Re}\left\{ \overline{\phi} \, \exp(-\rmi \omega t) \right\}$, but we omit reference to this henceforth for brevity. Due to the symmetries of  the full array problem, we consider Helmholtz's equation in  the     fundamental unit cell $\overline{\Omega}$ containing a single resonator and satisfying Bloch conditions   between adjacent cells (defined below). Here,  $(\overline{x},\overline{y})$ represents dimensional Cartesian coordinates,    $k  = \omega \sqrt{  \rho / B} $   the wave number, $\omega$   the angular frequency,     $B$ the Bulk modulus, and      $\rho$   the  mass density of the surrounding acoustic medium. For future reference, we also denote the dimensional Bloch vector by $(\overline{k}_{\rmB x},\overline{k}_{\rmB y})$, and we note that all Cartesian dimensional quantities   possess an overbar, along with $\overline{\phi}(\overline{x},\overline{y})$, but that the remaining  quantities do not (i.e., $\rho$, $\omega$, $B$, and $k$).  

 In order to reduce the number of parameters, and to better understand the mathematical treatment to follow, we    non-dimensionalise as
\begin{equation}
\label{eq:nondimk}
\overline{x} = x/k, \quad \overline{y} = y/k, \quad   \overline{b} = b/k, \quad \overline{d} = d/k, \quad \overline{\ell} = \ell/k,
\end{equation}
to obtain the governing equations for our problem inside the unit cell,   shown in Fig.~\ref{fig:resonatorarray}, in the form    
\begin{subequations}
\label{eq:nondimform}
\begin{align}
\label{eq:helmholtz}
(\partial_x^2+\partial_y^2 +1) \phi &= 0,  \\
\label{eq:neumann}
\frac{\partial\phi}{\partial r}  \bigg|_{S} &= 0, \\
\label{eq:bloch}
\phi(x + md,y+nd) &= \phi(x,y) \, \rme^{\rmi (k_{\rmB x} md + k_{\rmB y} nd )},
\end{align}
\end{subequations}
where  $\overline{\phi}(\overline{x},\overline{y}) = \phi(x,y)$,  we define   the potential $\phi(x,y) = \overline{\phi}(\overline{x}/k,\overline{y}/k)$, and we represent the infinitesimally thin cylinder with an aperture by
\begin{equation}
S = (b \cos \theta, b \sin \theta) \quad \mbox{with} \quad    \theta \in (\theta_0+\theta_\mathrm{ap},2\pi-\theta_\mathrm{ap}+\theta_0).
\end{equation}  The definition for $S$  prescribes an infinitesimally thin resonator of radius $b$ with an aperture centred at $\theta_0$ and a half-width angle of $\theta_\mathrm{ap} = \ell/b$ (i.e., a gap with total arc length $2\ell = 2 k \overline{\ell}$). For the Bloch condition \eqref{eq:bloch} we   define the integers $m,n\in\mathbb{Z}$ and lattice period for a square array $d$.
In this work we   treat half the arc length for the aperture as the small parameter $\varepsilon =\ell$,  as this is the appropriate regime for resonance, and we begin by considering the   problem local to the aperture to commence our asymptotic solution.

\section{Inner problem formulation}
\label{chap:innerthin}
As outlined in \cite{crighton1992modern,bender2013advanced,cotterill2015time}, solutions obtained using matched asymptotic methods require both an inner and an outer solution, in addition to a rigorous matching rule. In general, the inner solution describes the   near field (i.e., close to a boundary or object), and the outer solution describes the behaviour in the far field (i.e., far away from the boundary or object) \cite{crighton1992modern}.  For our   problem, the curvature of the resonator wall boundary is locally zero as we focus in on the aperture, and so, the walls may be regarded as flat (i.e., we take the asymptotic limit as the radius of the cylinder is long relative to the aperture size). This idea of vanishing local curvature is equivalent to the concept of a plane wave, which formally corresponds to a source point placed at infinity. 

\begin{figure}[t]
\centering
 \subfloat[Subfigure 1 list of figures text][]{
\includegraphics[width=0.41\textwidth]{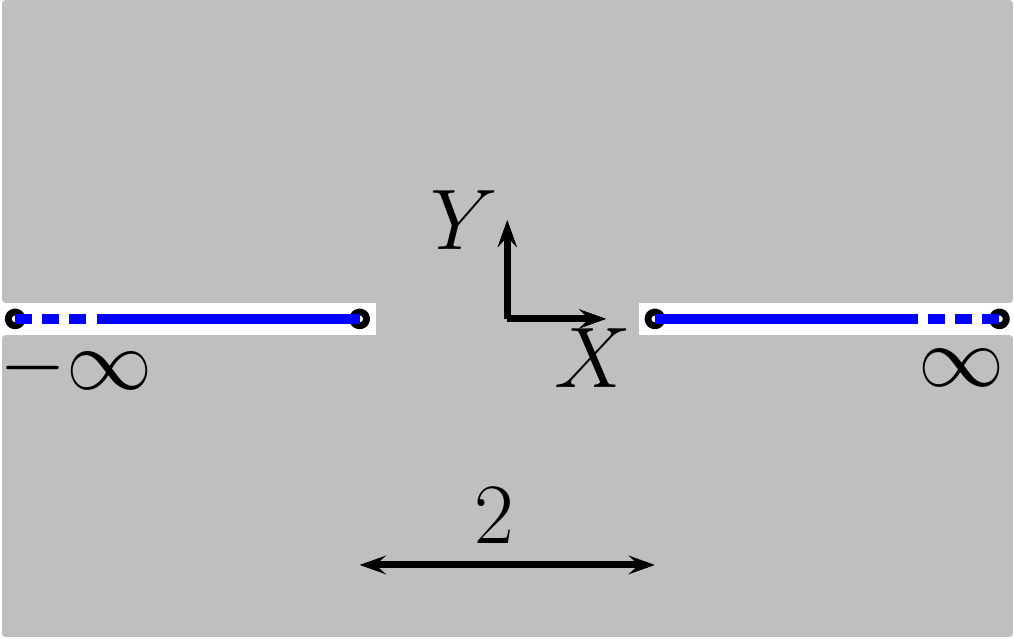}
\label{fig:innsubfig1} }\hspace{10mm} 
\subfloat[Subfigure 2 list of figures text][]{
\includegraphics[width=0.335\textwidth]{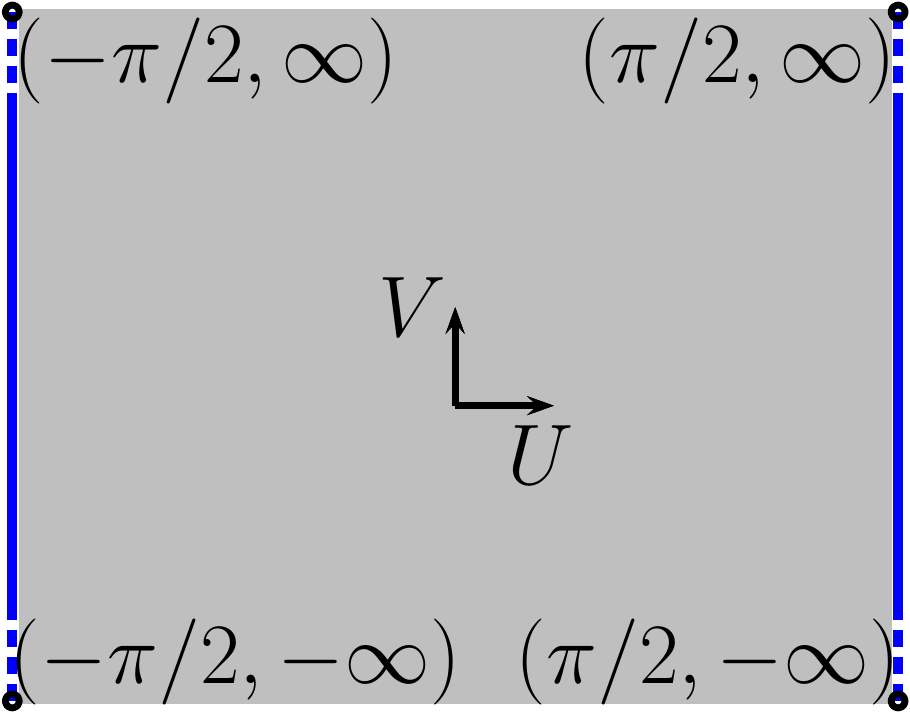}
\label{fig:innsubfig2}}

\caption{Inner problem domain comprising  an infinitely extending screen with resonator mouth   of length $2$ in terms of the inner coordinates $(X,Y)$; (b) equivalent representation obtained via the mapping $W = \arcsin(Z)$ in terms of transformed inner coordinates $(U,V)$.  \label{fig:schemorig}}
  \end{figure}

As a first step we rotate and translate the array via   $(\tilde{x},\tilde{y})\mapsto (x\sin\theta_0 - y\cos\theta_0,x\cos\theta_0 + y\sin\theta_0 - b)$ so that the aperture in the fundamental cell is centered about the origin. Subsequently we introduce   the inner scaling
\begin{equation}
\label{eq:innerexp}
X = \tilde{x}/\varepsilon, \quad \mbox{and} \quad Y = \tilde{y}/\varepsilon,
\end{equation}
 as well as the regular expansion $\phi  = \sum_{m=0}^{\infty}\varepsilon^m\Phi_m(X,Y)$.  
Substituting the scaling \eqref{eq:innerexp} and expansion into the Helmholtz equation \eqref{eq:helmholtz} and Neumann condition \eqref{eq:neumann},    we obtain the leading-order inner problem   given by
\begin{subequations}
\begin{align}
(\partial_X^2 + \partial_Y^2)  \Phi = 0, \quad \mbox{ for } X  \in \mathbb{R}^2 \backslash S^\mathrm{in}, \\
\partial_{Y} \Phi  = 0, \quad \mbox{ for } X  \in S^\mathrm{in},
\end{align}
\end{subequations}
where   $S^\mathrm{in} = \left\{(X ,Y ): Y =0, |X|\geq 1 \right\}$, i.e., the geometry looks locally planar as shown in Fig.~\ref{fig:schemorig}, and we   omit the subscript for $\Phi_0$ for clarity. Next we introduce the  mapping
$W = \arcsin(Z)$
where $Z = X + \rmi Y = R \, \mathrm{exp}(\rmi \Theta)$ and $W = U+\rmi V$, which transfers the problem of solving Laplace's equation in $\mathbb{R}^2\backslash S^\mathrm{in}$  to solving Laplace's equation in an infinitely extending strip, as shown in Figure \ref{fig:innsubfig2}, and described by
\begin{subequations}
\begin{align}
( \partial_U^2 + \partial_V^2)  \Phi_\rmD = 0, \quad \mbox{ for } U  \in D, \\
\partial_{U} \Phi_\rmD  = 0, \quad \mbox{ for } U  =\pm \pi/2,
\end{align}
\end{subequations}
where   $D =  \left\{(U ,V ): V \in (-\infty,\infty), |U|\leq \pi/2 \right\}$. The appropriate solution   is given by
\begin{equation}
 \Phi_\mathrm{D}= C_1 \mathrm{Re}(\rmi W) + C_2,
\end{equation}
where $C_j$ are as yet unknown, and we will see in the following sections why this form is the appropriate solution for matching. Subsequently,  the solution in the original domain    follows as
\begin{equation}
\label{eq:PhiCjfinal}
\Phi=     C_1 \mathrm{Re}(\rmi \arcsin(Z)) + C_2 = C_1 \mathrm{Re}\left\{ \log(\rmi Z + \sqrt{1-Z^2}) \right\} + C_2,
\end{equation}
where we define $\sqrt{1-Z^2} = \rmi \sqrt{Z^2 - 1}$ (i.e.,   the positive branch).

\subsection{Limiting behaviour of inner solution as {\large $R = |X^2+Y^2|^{1/2}\rightarrow \infty$}}
We now require the field $\Phi$ as $R\rightarrow \infty$ in both the lower- and upper-half planes.
To ensure single-valuedness 
we     introduce the   double-angle representation 
\begin{subequations}
\begin{equation}
\label{eq:z2m1def}
\sqrt{Z^2 - 1} =   \sqrt{|Z^2 - 1|}\rme^{\rmi (\Theta_1 +  \Theta_2)/2},
\end{equation} \sloppy
over the cut plane   $Z \in \mathbb{C}\backslash B_\mathrm{C}$ where $Z - 1  = R_1 \mathrm{exp}(\rmi \Theta_1)$ and  $Z + 1  =  R_2 \mathrm{exp}(\rmi \Theta_2)$  for  $ \Theta_1 \in (-2\pi,0)$ and $\Theta_2 \in (-\pi,\pi)$, with  $B_\mathrm{C} = \left\{ (X,Y): X \in (\infty,-1) \cup (1,\infty) \times Y=0\right\}$ denoting the  branch cuts. Thus, if we   proceed to infinity in the upper-half plane (i.e.,   $\Theta_1 \rightarrow -3\pi/2$ and $\Theta_2 \rightarrow \pi/2$) and in the lower-half plane  (i.e., $\Theta_1 \rightarrow -\pi/2$ and $\Theta_2 \rightarrow -\pi/2$) we obtain
\begin{equation}
\lim_{Z \rightarrow \rmi \infty}\sqrt{Z^2 - 1} \approx
  -Z +\frac{1}{2Z} + O(Z^{-3}),
  \quad \mbox{ and } \quad
  \lim_{Z \rightarrow -\rmi \infty}\sqrt{Z^2 - 1} \approx
  Z   + O(Z^{-1}).
\end{equation}
\end{subequations}
Accordingly for the inner solution has, from \eqref{eq:PhiCjfinal}, the asymptotic form
\begin{equation}
\label{eq:rhsasy}
\left( \lim_{R\rightarrow \infty}\Phi \right) \bigg|_{R = \tilde{r}/\varepsilon}\sim 
\begin{cases}
 -C_1 \left[ \log(\tilde{r}) - \log\left(\dfrac{\varepsilon}{2}\right) \right]+ C_2,   & Z \in \mathbb{C}^\rmU, \vspace{1mm} \\
 \phantom{-}C_1 \left[ \log( \tilde{r}) -\log\left(\dfrac{\varepsilon}{2}\right) \right] + C_2,  & Z \in \mathbb{C}^\rmL,
\end{cases}
\end{equation}
  where $\mathbb{C}^{\rmU}$ and $\mathbb{C}^{\rmL}$ denote the upper- and lower-half segments of the complex plane, respectively, and where we re-express the solution  with respect to the original outer coordinate frame.   We now proceed to the outer problem for our resonator array.

\section{Outer problem formulation}
\label{chap:outthin}

The  leading-order system for the outer problem is   obtained  by   taking the limit $\varepsilon \rightarrow 0$ directly in the formulation \eqref{eq:nondimform} above to obtain the   system
\begin{subequations}
\label{eq:nondimform_outer}
\begin{align}
\label{eq:helmholtz_outer}
(\partial_x^2 + \partial_y^2 +1) \phi  &= 0,  \\
\label{eq:neumann_outer}
\frac{\partial\phi }{\partial r}  \bigg|_{S_\mathrm{out}} &= S_0, \\
\label{eq:bloch_outer}
\phi(x + md,y+nd) &= \phi(x,y) \, \rme^{\rmi (k_{\rmB x} md +  k_{\rmB y} nd  ) },
\end{align}
\end{subequations}
which is defined inside the fundamental unit cell $\Omega_\mathrm{out}$. This is itself almost identical to the original unit cell $\Omega$ in the system  \eqref{eq:nondimform} except that the resonator is almost closed, i.e.,  it is  defined by $S_\mathrm{out} = (r \cos \theta, r \sin \theta)$, with $r=b$ and $\theta \in (0,2\pi)\backslash \theta_0$, with the aperture acting as a yet to be determined point source $S_0$ at $\theta = \theta_0$. For the outer array problem defined above   \eqref{eq:nondimform_outer} we now decompose the unit cell into two domains and  consider a solution outside the resonator (the outer exterior solution $\phi_\mathrm{ext}$) and inside the resonator (the outer interior solution $\phi_\mathrm{int}$). 

\subsection{Outer exterior ansatz}
\label{sec:outerextthin}
In the  region exterior to the resonator, but inside the fundamental unit cell, we pose the ansatz 
\begin{equation}
\label{eq:basicansatzHr}
\phi_\mathrm{ext} = A \rmH_0^{(1)}(\widetilde{r}) + \sum_{n=-\infty}^{\infty} \left\{ a_n \rmJ_n(r)  +   b_n \rmY_n(r)  \right\} \rme^{\rmi n \theta}  ,
\end{equation}
  where $A$, $a_n$, and $b_n$ are as yet unknown,   $\widetilde{r}^2 = r^2 + b^2 -2 r b \cos(\theta - \theta_0)$,   $\rmJ_n(z)$ and $\rmY_n(z)$ denote Bessel functions of the first and second kind, respectively, and $\rmH_n^{(1)}(z)$ represent   Hankel functions of the first kind. We remark that  the impact of periodicity will be incorporated later in Section \ref{chap:outthin}\ref{chap:latticeray}.    Next, we   express the Neumann boundary condition  \eqref{eq:neumann_outer} as
\begin{equation}
\label{eq:neumannouter}
\frac{\partial\phi_\mathrm{ext} }{\partial r}  \bigg|_{r=b} = S_0 =  \frac{C}{b} \delta(\theta - \theta_0) =   \frac{C}{2 \pi b} \sum_{n=-\infty}^{\infty}  \rme^{\rmi n (\theta - \theta_0)},
\end{equation}
where   $C$ is unknown. The relationship between $C$ and $A$ is determined by applying Graf's addition theorem \cite[Eq. (8.530)]{gradshteyn2014table}
\begin{equation}
\label{eq:gadh0}
   \rmH_0^{(1)}(\widetilde{r})
   =\begin{cases}
   \sum\limits_{n=-\infty}^{\infty} \rmJ_n(b) \rmH_n^{(1)}(r)\rme^{\rmi n (\theta - \theta_0)}, \quad r>b,  \\
   \sum\limits_{n=-\infty}^{\infty} \rmJ_n(r) \rmH_n^{(1)}(b)\rme^{\rmi n (\theta - \theta_0)}, \quad r<b ,
   \end{cases}
\end{equation}
 and taking the limit  $n \rightarrow \infty$. By  matching     the Dirac delta   singularity in \eqref{eq:neumannouter}   with the logarithmic singularity in the solution ansatz at $(b,\theta_0)$ in \eqref{eq:basicansatzHr}, we obtain the necessary form of the boundary condition $ C = 2 \rmi A$. Subsequently, after imposing the updated Neumann condition we obtain
\begin{equation}
\label{eq:genformneumannsatA}
\phi_\mathrm{ext} = A \rmH_0^{(1)}(\widetilde{r}) - \sum_{n=-\infty}^{\infty}
 \left\{     \frac{A Q_n}{2 \rmJ_n^\prime(b)}   \rme^{-\rmi n    \theta_0 }  + \frac{\rmY_n^\prime(b)}{\rmJ_n^\prime(b)} b_n \right\}   \rmJ_n(r)  \rme^{\rmi n \theta  } +
\sum_{n=-\infty}^{\infty}b_n    \rmY_n(r)     \rme^{\rmi n \theta} ,
\end{equation}
where
\begin{equation}
   Q_m =    	\rmJ_m(b)\rmH_m^{(1)\prime}(b) + \rmJ_m^\prime(b) \rmH_m^{(1)}(b) .
\end{equation}
Note that the constants $b_n$ and $A$ are as yet unknown, but that the form of $\phi_\mathrm{ext}$ is   prescribed.

 \subsection{Outer interior ansatz} \label{sec:outintthin}
Next we construct a  corresponding form of the outer solution inside the resonator   following an identical procedure to the above. Hence, we write
\begin{equation}
\phi_\mathrm{int} = B \, \rmH_0^{(1)}(\widetilde{r}) + \sum_{n=-\infty}^{\infty} f_n \rmJ_n(r) \rme^{\rmi n \theta},
\end{equation}
  and consider the Neumann boundary condition 
\begin{equation}
\frac{\partial  \phi_\mathrm{int}}{\partial r}\bigg|_{r=b} = \dfrac{D}{b}\delta(\theta-\theta_0),
\end{equation}
where $D$ is once more unknown.  Imposing the Neumann condition above, using Graf's addition theorem  \eqref{eq:gadh0}, and considering the limit $n\rightarrow \infty$ we find that $D = -2\rmi B$ on matching divergent terms. Imposing the updated Neumann condition yields  
\begin{equation}
\label{eq:genformneumannsatB}
\phi_\mathrm{int} = B\rmH_0^{(1)}(\widetilde{r}) - \frac{B}{2}\sum_{n=-\infty}^{\infty} \frac{Q_n}{\rmJ_n^\prime(b)} \rmJ_n(r) \rme^{\rmi n (\theta-\theta_0)},
\end{equation}
where $B$ is unknown. We can now take the outer solutions in the exterior  \eqref{eq:genformneumannsatA} and interior \eqref{eq:genformneumannsatB} domains, and determine their   asymptotic forms as we approach the aperture
\begin{equation}
\label{eq:outerasympsbothdir}
\lim_{\theta \rightarrow \theta_0} \lim_{r\rightarrow b} \phi  \sim
\begin{cases}
 \dfrac{2\rmi A}{\pi} \left[ \gamma_\rme - \dfrac{\rmi \pi}{2} + \log\left(\dfrac{\tilde{r}}{2}\right) \right]  + \sum\limits_{n=-\infty}^{\infty}b_n    \rmY_n(b)     \rme^{\rmi n \theta_0}
  \\ 
 \hspace{40mm}- 
\sum\limits_{n=-\infty}^{\infty} \left\{     \dfrac{A Q_n}{2  }    + b_n \rmY_n^\prime(b)    \rme^{ \rmi n    \theta_0 } \right\}   \dfrac{\rmJ_n(b)}{\rmJ_n^\prime(b)}  
 , &  r \downarrow b, \\
\dfrac{2\rmi B}{\pi} \left[ \gamma_\rme - \dfrac{\rmi \pi}{2} + \log\left(\dfrac{\tilde{r}}{2}\right) \right] - \dfrac{B}{2}\sum\limits_{n=-\infty}^{\infty} \dfrac{Q_n}{\rmJ_n^\prime(b)} \rmJ_n(b), \phantom{\bigg|^b}	& r \uparrow b ,
\end{cases}
\end{equation}
where $\gamma_\rme$ denotes the Euler--Mascheroni constant. Having determined partial solutions (up to an infinite set of constants) for both the inner and outer solutions, and their asymptotic representations near the aperture, we now proceed to   asymptotic matching. 
\subsection{Matched asymptotics procedure with partial solutions}
\label{eq:matchedasy}
The unknown terms $A$, $B$, $C_1$ and $C_2$  in the above are obtained by matching terms (to leading order) from the inner and outer solution representations in the following limit \cite{crighton1992modern,bender2013advanced}
\begin{equation}
\lim_{\theta \rightarrow \theta_0} \lim_{r\rightarrow b} \phi  = \left( \lim_{R\rightarrow \infty} \Phi\right)\bigg|_{R = r/\varepsilon},  
\end{equation}
where   the asymptotic forms are given above in \eqref{eq:outerasympsbothdir} and \eqref{eq:rhsasy}. Specifically, we match the inner  and outer solutions, in the interior/lower and exterior/upper regions, and then  compare logarithmic and non-logarithmic terms to obtain a   system of   equations. These yield  $B=-A$   and 
\begin{equation}
\label{eq:Abnrel}
A = 
  \dfrac{2}{\pi b h_\varepsilon}\sum\limits_{n=-\infty}^{\infty}  \dfrac{b_n}{\rmJ_n^\prime(b)} \rme^{ \rmi n    \theta_0 }    ,
\end{equation} 
where
\begin{equation}
\label{eq:hepsfull}
h_\varepsilon =\dfrac{4\rmi  }{\pi} \left[ \gamma_\rme - \dfrac{\rmi \pi}{2} + \log\left(\dfrac{\varepsilon}{4}\right) \right] - \sum\limits_{n=-\infty}^{\infty} \dfrac{Q_n  \rmJ_n(b)}{\rmJ_n^\prime(b)}  .
\end{equation}

\subsection{Lattice contributions and asymptotic resonator system}
\label{chap:latticeray}
The final step in our derivation of an eigenvalue problem for the resonant array  involves returning to the exterior solution ansatz \eqref{eq:genformneumannsatA} and applying Graf's addition theorem \eqref{eq:gadh0} to obtain
\begin{multline}
\label{eq:genformneumannsat3}
\phi_\mathrm{ext} =  \sum\limits_{n=-\infty}^{\infty} \left[A \rmJ_n(b) \rme^{-\rmi n  \theta_0}  - \frac{A}{2} \frac{Q_n}{\rmJ_n^\prime(b)} \rme^{-\rmi n  \theta_0}  - \frac{\rmY_n^\prime(b)}{\rmJ_n^\prime(b)}   b_n \right] \rmJ_n(r) \rme^{\rmi n  \theta} \\
+  \sum\limits_{n=-\infty}^{\infty} \left[ \rmi A  \rmJ_n(b) \rme^{-\rmi n  \theta_0}  + b_n\right] \rmY_n(r)\rme^{\rmi n  \theta  } 
 =    \sum\limits_{n=-\infty}^{\infty} \left\{ c_n \rmJ_n(r)+ d_n \rmY_n(r) \right\}\rme^{\rmi n  \theta},
\end{multline}
where the $c_n$ and $d_n$ coefficients are related through the dynamic Rayleigh identity \cite[Eq. (3.119)]{movchan2002asymptotic}
\begin{equation}
\label{eq:rayid}
c_n = \sum_{m=-\infty}^{\infty} (-1)^{m+n}S_{m-n}^\rmY(\bfk_\rmB) d_m,
\end{equation}
which   follows   from an application of Green's second identity inside the unit cell. The Rayleigh identity incorporates multiple scattering contributions from neighbouring cells by imposing   the Bloch conditions \eqref{eq:bloch_outer}. Expressions for the lattice sums $S_m^\rmY$ are given in Appendix \ref{chap:SlY} for reference.

At this point, we remark that we possess an identity  relating $c_n$ and $d_n$ in \eqref{eq:rayid},   expressions for $c_n$ and $d_n$ in terms of $A$ and $b_n$ in \eqref{eq:genformneumannsat3}, and a relation between $A$ and $b_n$ from the matched asymptotics procedure \eqref{eq:Abnrel}. Merging all of these expressions we   obtain the    eigenvalue problem
\begin{multline}
\label{eq:dispeqsystemgn}
 \dfrac{ \rmi    }{\pi b h_\varepsilon} \left(\sum\limits_{q=-\infty}^{\infty} g_q  \right) \left[    \frac{\rmJ_n^\prime(b)\rmY_n(b) + \rmY_n^\prime(b)   \rmJ_n(b)}{\rmJ_n^\prime(b) \rmY_n^\prime(b)}  \right] + g_n \\
 +      \sum_{m=-\infty}^{\infty} (-1)^{m+n}S_{m-n}^\rmY(\bfk_\rmB) \frac{\rmJ_m^\prime(b) }{\rmY_n^\prime(b)} \rme^{-\rmi(m-n)\theta_0}	g_m   \\
+  \dfrac{2\rmi}{\pi b h_\varepsilon}  \left(\sum\limits_{q=-\infty}^{\infty}g_q  \right)   \left( \sum_{m=-\infty}^{\infty} (-1)^{n+m}S_{m-n}^\rmY(\bfk_\rmB) \frac{\rmJ_m(b)}{\rmY_n^\prime(b)} \rme^{-\rmi (m-n)  \theta_0} \right) = 0,
\end{multline} 
which must be satisfied for all $n\in \mathbb{Z}$. For reference, the representation \eqref{eq:dispeqsystemgn} is obtained after introducing the scaling 
\begin{equation}
b_n = \rmJ_n^\prime(b) g_n \rme^{-\rmi n \theta_0},
\end{equation}
 and after multiplying  by the factor $\exp(\rmi n \theta_0)/\rmY_n^\prime(b)$. We remark that upon closing the gap $\varepsilon \rightarrow 0$ then $h_\varepsilon \rightarrow -\rmi \infty$ and we recover the conventional dispersion equation system for an array of homogeneous Neumann cylindrical inclusions \cite[Eq. (3.158)]{movchan2002asymptotic}. Next, for  numerical and analytical purposes, we require that the infinite dimensional system \eqref{eq:dispeqsystemgn}, and all sums contained therein, are truncated in order to obtain a finite-dimensional system; imposing a vanishing determinant condition then yields the dispersion equation for that truncation (denoted by the truncation level $L$), where the accuracy is generally improved  as we truncate to   higher orders. For reference,    care must be taken for large $L$ as accurate band diagrams may be constructed but inaccurate modal fields  may arise  (i.e., from \eqref{eq:genformneumannsatA} and \eqref{eq:genformneumannsatB}) as errors in the asymptotic approximations dominate.

It is well-known that for periodic problems involving cylinders with Neumann boundary conditions, a   monopole truncation      is  unable  to accurately describe the   eigenstate at low frequencies. As such, we proceed to the next section by considering a dipolar truncated system.

\subsection{Leading-order approximation to the dispersion equation}
\label{eq:leadingordapprox}
Considering the system \eqref{eq:dispeqsystemgn} we now truncate all sums, and consider  all orders, to  within a dipole approximation $L=1$ (i.e., keeping terms $n=-1, 0, 1$) to construct the dipole system. We then evaluate Taylor series in the small $b$ (long wavelength) limit to obtain  the leading-order system $\bfA_0 \bfg = \boldsymbol{0}$ given by 
\begin{equation}
\label{eq:systemdipole}
\left[
\begin{array}{c|c|c}
1+\dfrac{1}{4} \pi b^2 S_0^\rmY 
&
\phantom{\bigg|^{1}}
 -\dfrac{1}{h_\varepsilon}\rmi b S_1^{\rmY\ast} \rme^{\rmi \theta_0} +\dfrac{1}{4} \pi b^3  S_1^{\rmY \ast} \rme^{\rmi \theta_0}   
\phantom{\bigg|^{1}}
& -\dfrac{1}{4}\pi b^2 S_2^{\rmY \ast}\rme^{2\rmi  \theta_0  }   \\ 
\hline 
 -\dfrac{1}{4} \pi b S_1^\rmY  \rme^{-\rmi \theta_0}
&
\phantom{\bigg|^{1}}
  -\dfrac{2\rmi}{\pi b^2 h_\varepsilon}  + \dfrac{\rmi S_0^\rmY}{h_\varepsilon} + 1  -\dfrac{1}{4}\pi b^2 S_0^\rmY 
\phantom{\bigg|^{1}}
&  \dfrac{1}{4} \pi b  S_1^{\rmY \ast} \rme^{\rmi \theta_0} \\ \hline
-\dfrac{1}{4} \pi b^2 S_2^\rmY  \rme^{-2 \rmi \theta_0}
& 
\phantom{\bigg|^{1}}
\dfrac{1}{h_\varepsilon}\rmi b S_1^{\rmY} \rme^{-\rmi \theta_0} -\dfrac{1}{4} \pi b^3  S_1^{\rmY } \rme^{-\rmi \theta_0}  
 \phantom{\bigg|^{1}}
& 1+\dfrac{1}{4} \pi b^2 S_0^\rmY 
\end{array}
\right]
\left[ 
\begin{array}{c}
g_1 \\
g_0 \\
g_{-1}
\end{array} 
\right]
=
\left[ 
\begin{array}{c}
0 \\
0 \\
0
\end{array} 
\right],
\end{equation}
 and where we have   made use of the   lattice sum asymptotic forms   \cite[Eq. (3.132)--(3.134)]{movchan2002asymptotic} in the long  wavelength  and low frequency limits
\begin{subequations}
\label{eq:slYgall}
\begin{equation}
\lim_{ {k}_\rmB\rightarrow \Gamma}\lim_{k\rightarrow 0}\left\{ S_0^\rmY,S_1^\rmY,S_2^\rmY\right\}  \sim \left\{-\frac{4}{d^2} \frac{1}{k_\rmB^2 -1}, -\frac{4\rmi k_\rmB }{d^2} \frac{\rme^{\rmi \theta_\rmB}}{k_\rmB^2 -1}, \frac{4k_\rmB^2}{d^2} \frac{ \rme^{2\rmi \theta_\rmB}}{k_\rmB^2 -1}\right\}
\end{equation}
\end{subequations}
with $(k_\rmB,\theta_\rmB)$ representing the polar form of the Bloch vector $\bfk_{\rmB}$ and $\ast$ denoting the complex conjugate operation. In the system above, we have also made use of the   dominant balance scaling $d = O(b)$, to avoid implicitly examining the vanishing filling fraction $f = \pi b^2/d^2$    limit  as $b\rightarrow 0$, and also used   the dominant balance scaling $\log(\varepsilon/(2b)) = O(b^{-2})$   appearing in 
\begin{subequations}
\begin{equation}
\label{eq:hepsfeps}
\lim_{b\rightarrow 0} h_\varepsilon \sim \frac{4\rmi}{\pi b^2} \left[1 - \frac{b^2}{8}  + b^2\log\left( \frac{\varepsilon}{2b}\right) \right] = \frac{4\rmi}{\pi b^2} f_\varepsilon.
\end{equation}
Next we introduce the substitutions
\begin{equation}
S_0^\rmY = \frac{4}{\pi b^2} A_0, \quad S_1^\rmY = \frac{4\rmi}{\pi b^2} A_1 \rme^{\rmi \theta_\rmB}, \quad S_2^\rmY = \frac{4}{\pi b^2} A_2 \rme^{2\rmi \theta_\rmB} ,
\end{equation}
\end{subequations}
where $A_{0}$, $A_1$, and $A_2$ are strictly real, and  evaluate the determinant of the system \eqref{eq:systemdipole} to obtain the leading-order   dispersion equation
\begin{equation}
\label{eq:dipdispeq}
(1 + A_0 + A_2) \left\{ \left(1-\frac{1}{f_\varepsilon}\right) \left[A_0 (1 + A_0-A_2)-2 A_1^2\right]-\left(1-\frac{1}{2 f_\varepsilon}\right) (1 + A_0-A_2) \right\}=0,
\end{equation}
and so after returning to the  forms for $S_m^\rmY$ in \eqref{eq:slYgall} once more we obtain the lowest-order approximation for the dispersion equation of the first band in the form
\begin{equation}
\label{eq:dispeqdipoleleading}
k_\rmB^2 = \frac{1+f}{1-f} \left(1 -  \frac{   2 f (1 - f_\varepsilon )}{1 - 2 f_\varepsilon} \right), \quad \mbox{where we repeat that} \quad f=\frac{\pi b^2}{d^2}.
\end{equation} 
In the limit of vanishing aperture we have that $f_\varepsilon \rightarrow \infty$ and subsequently we recover the  classical result for an array of Neumann cylinders \cite[Eq. (3.158)]{movchan2002asymptotic} 
 \begin{equation}
 \label{eq:disprelneucyl}
k_\rmB^2 =  1+f .
\end{equation} 
 Thus, by specifying purely geometric parameters, namely the radius $\bar{b}$, half-angle $\theta_\mathrm{ap}$, and filling fraction $f$, as well as the Bloch wave vector $\bar{\bfk}_\rmB$, it is then possible to solve for $k$ in \eqref{eq:correctedsls} and obtain the leading-order approximation to the first band surface   over the entire Brillouin zone. Note that the absence of  the central angle $\theta_0$ in the above means that the leading-order approximation is unable to  describe the low-frequency anisotropy present in the  first band. For this reason, we proceed to a first-order correction, but first discuss some of the features of the approximation \eqref{eq:dispeqdipoleleading}. For example,   by substituting the leading-order behaviour
 \begin{equation}
 \label{eq:fepsasympt}
f_\varepsilon =  \left[1 - \frac{b^2}{8} + b^2\log \left( \frac{\varepsilon}{2b} \right)\right] \sim 1 + b^2\log\left(\frac{\theta_\mathrm{ap}}{2} \right),
\end{equation}
 into  the dispersion equation \eqref{eq:dispeqdipoleleading},  we obtain the result   presented in   Llewellyn--Smith \cite{llewellyn2010split}, which we correct for an errant sign error to:
\begin{equation}
\label{eq:correctedsls}
2 b^2 \log \left(\frac{2}{\theta_\mathrm{ap}}\right) = \frac{ k_\rmB^2 - (1+f)/(1-f)  }{k_\rmB^2 - (1+f)  }.
\end{equation}
Following the discussion in Llewellyn--Smith \cite{llewellyn2010split},  under the limit of vanishing aperture $\theta_\mathrm{ap} \rightarrow 0$ the representation \eqref{eq:correctedsls}  returns the classical result for an array of Neumann cylinders as given in \eqref{eq:disprelneucyl}. Likewise in the low-frequency limit $k \rightarrow 0$ we obtain a lowest-order dispersion relation for our  Helmholtz resonator array   \cite{llewellyn2010split} in the form
\begin{equation}
 k_\rmB^2 = \frac{1+f}{1-f}, \quad \mbox{ or } \quad \omega =  \sqrt{ \frac{B}{\rho} } \sqrt{\frac{1-f}{1+f}} \,\bar{k}_\rmB,
\end{equation}
however numerical investigations show   this leading-order result     to be   accurate       along one high-symmetry direction alone.
As a final remark on the leading-order dispersion equation \eqref{eq:dispeqdipoleleading}, we note that  although it  is unable to correctly describe the first band,  it is able to approximate  the cut-off frequency of the first band to within moderate accuracy (i.e., the maximum eigenfrequency of the first band but not necessarily the lower bound on the first band gap). This is done using  the vanishing denominator condition  in \eqref{eq:dispeqdipoleleading} to obtain
\begin{equation}
\label{eq:maxeval1stband}
  k_\mathrm{max} \approx \frac{2}{\bar{b}}\sqrt{\frac{1}{1 + 8 \log\left(2/\theta_\mathrm{ap}\right)}}, \quad \mbox{ or } \quad \omega_\mathrm{max} \approx \frac{2}{\bar{b}}  \sqrt{\frac{B}{\rho}}\sqrt{\frac{1}{1 + 8 \log\left(2/\theta_\mathrm{ap}\right)}}.
\end{equation}
The above expression may also be used to determine the configuration of the resonator within the unit cell, for example, if we seek a resonance in air $c_p = \sqrt{B/\rho} = 343.21$ m/s at the frequency $f_\mathrm{max} = 60$ Hz with the (arbitrary) design restriction of the half aperture length being $\bar{\ell} = 0.01$ m = 1 cm, then we require a radius of $\bar{b} = 0.312$ m = 31.2 cm.
 
\subsection{First-order correction to the dispersion equation}
\label{chap:correction}
The first-order correction to the system \eqref{eq:dispeqsystemgn} within a dipolar truncation takes the form
\begin{subequations}
\begin{equation}
\bfB \bfg= \boldsymbol{0},
\end{equation} 
where $\bfB = \bfA_0 + \bfA_1$, with $\bfA_0$ given in \eqref{eq:systemdipole}, and
\begin{equation}
\label{eq:defmatA1}
\bfA_1 = 
\left[
\begin{array}{c|c|c}
- \dfrac{\rmi b }{h_\varepsilon}   S_1^{\rmY \ast} \rme^{\rmi \theta_0}
&
\phantom{\bigg|^1}
  \dfrac{ \rmi b^2}{2 h_\varepsilon} \left(S_0^\rmY   -  S_2^{\rmY \ast}\rme^{ 2 \rmi \theta_0} \right)  
\phantom{\bigg|^1}
&
-\dfrac{\rmi b }{h_\varepsilon} S_1^{\rmY \ast} \rme^{\rmi \theta_0}  \\ \hline
-\dfrac{2\rmi}{\pi b^2 h_\varepsilon} + \dfrac{\rmi }{h_\varepsilon}S_0^\rmY   
&
\phantom{\bigg|^1}
  \dfrac{\rmi b}{2 h_\varepsilon} \left(    S_1^{\rmY \ast}\rme^{\rmi \theta_0} -   S_1^{\rmY}\rme^{-\rmi \theta_0} \right)   
\phantom{\bigg|^1}
&
-\dfrac{2\rmi}{\pi b^2 h_\varepsilon} + \dfrac{\rmi }{h_\varepsilon}S_0^\rmY   \\ \hline
\dfrac{\rmi b }{h_\varepsilon} S_1^{\rmY } \rme^{-\rmi \theta_0}  
&
\phantom{\bigg|^1}
  \dfrac{ \rmi b^2}{2 h_\varepsilon} \left(S_0^\rmY   -  S_2^{\rmY }\rme^{ -2 \rmi \theta_0} \right)  
\phantom{\bigg|^1}
&
\dfrac{\rmi b }{h_\varepsilon}   S_1^{\rmY} \rme^{-\rmi \theta_0}
\end{array}
\right].
\end{equation}
\end{subequations}
   Solving for $\det \bfB = 0$ we   obtain the principal result of this paper:
   \begin{subequations}
\begin{equation}
\label{eq:dispeqfirstorder}
k_\rmB^2 = \frac{(f+1) \left[b^2 f  (2 f-1) +   f_\varepsilon   (f+1) (2 f_\varepsilon f-2 f_\varepsilon-2 f+1)\right]}{b^2 f  \cos (2 \left[\theta_0-\theta_\rmB\right])+b^2 f^2+  f_\varepsilon (2 f_\varepsilon-1) \left(f^2-1\right)},
\end{equation}
which is an asymptotic dispersion equation implicitly describing the first spectral band surface. The  expression   \eqref{eq:dispeqfirstorder} above  is crucially able to capture   the low-frequency anisotropy (i.e., differing low-frequency slopes)  present  in the first spectral band.  In the closed aperture limit $f_\varepsilon \rightarrow \infty$ we recover the result for an array of perfect Neumann cylinders $k_\rmB^2 = 1+ f$ from   \eqref{eq:dispeqfirstorder} above. We now briefly discuss the features of the dispersion relation derived above; note the presence of all angular dependencies: $\theta_0$, $\theta_\mathrm{ap}$ (via $f_\varepsilon$), and  $\theta_\rmB$, and, that    the analogue to \eqref{eq:correctedsls} is considerably less compact as $f_\varepsilon^2$ terms are present. Rearranging \eqref{eq:dispeqfirstorder} above we obtain a low-frequency dispersion relation of the form $\omega  = c_p^\mathrm{eff}(\theta_\rmB,\omega) \, \bar{k}_\rmB$ where
\begin{equation}
\label{eq:dispeqfirstorderomega}
  c_p^\mathrm{eff}(\theta_\rmB,\omega) = \sqrt{ \frac{B}{\rho} }  \sqrt{\frac{b^2 f  \cos (2 \left[\theta_0-\theta_\rmB\right])+b^2 f^2+  f_\varepsilon (2 f_\varepsilon-1) \left(f^2-1\right)}{(f+1) \left[b^2 f  (2 f-1) +   f_\varepsilon   (f+1) (2 f_\varepsilon f-2 f_\varepsilon-2 f+1)\right]} }  ,
\end{equation}
\end{subequations}
in which the effective wave speed $c_p^\mathrm{eff}$  possesses dependence on  both the frequency and Bloch vector direction (i.e.,   exhibits both temporal and spatial dispersion).  By  differentiating \eqref{eq:dispeqfirstorder}    the group velocity   is obtained explicitly but is not included here for compactness. A detailed discussion of homogenisation methods     is made in Part II of this work. 

 \section{Numerical Results}
 \label{chap:numericals}
 
  \begin{figure}[t]
\centering
\subfloat[Subfigure 1 list of figures text][]{
\includegraphics[width=0.475\textwidth]{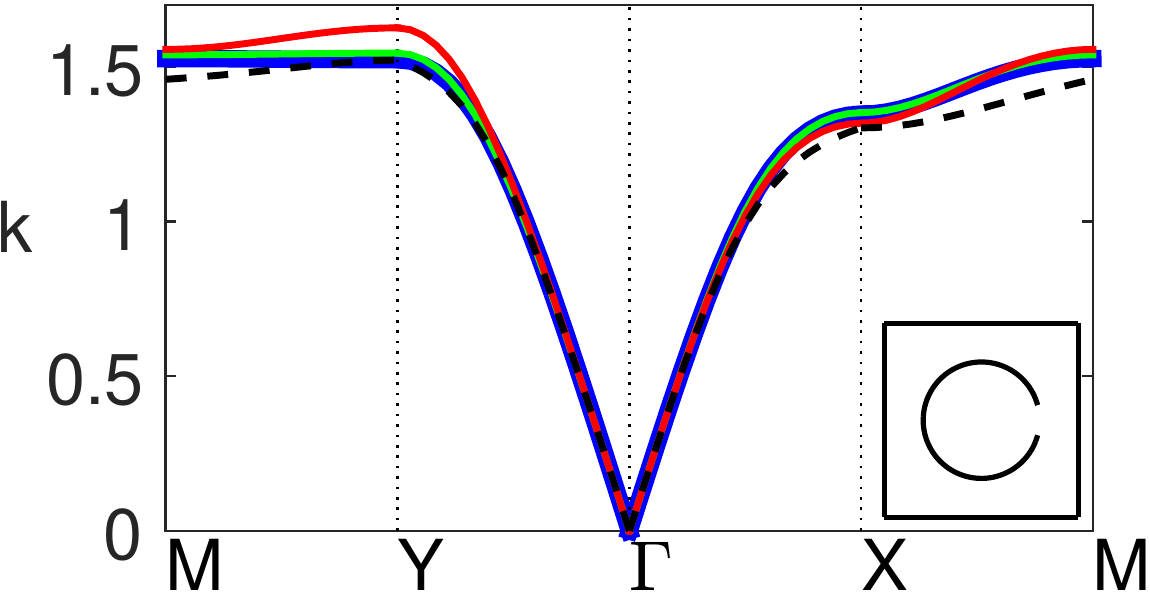}
\label{fig:comparefigssubfig1}}
\subfloat[Subfigure 2 list of figures text][]{
\includegraphics[width=0.475\textwidth]{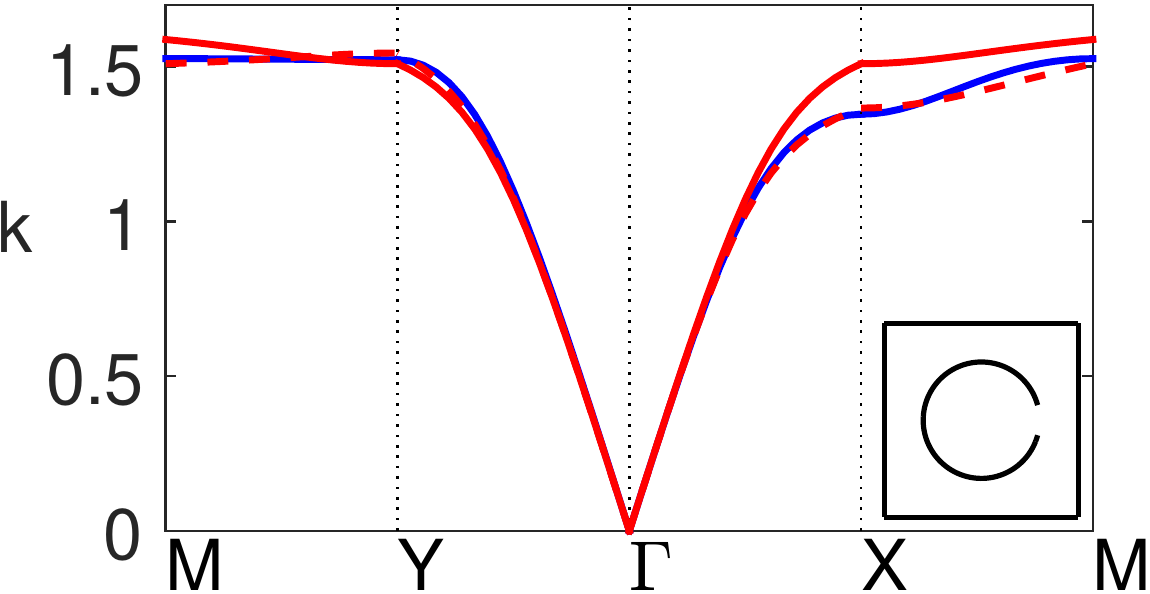}
\label{fig:comparefigssubfig2}} 
\caption{Band diagrams for a two-dimensional square array of thin-walled Helmholtz resonators comparing: \protect\subref{fig:comparefigssubfig1}   the   finite-element solution (blue line) with results from the system \eqref{eq:dispeqsystemgn} within a dipole truncation using either the convergent forms for $S_m^\rmY$   \eqref{eq:convergentSmY} (red line) or the asymptotic forms for $S_m^\rmY$   \eqref{eq:slYgall} (dashed black line), and results from \eqref{eq:dispeqsystemgn} within a quadrupolar truncation (green line);  \protect\subref{fig:comparefigssubfig2}   the full finite-element solution (blue line) with the symmetric lowest-order approximation \eqref{eq:dispeqdipoleleading} (solid red line) and the first-order correction \eqref{eq:dispeqfirstorder} (dashed red line). In both figures we specify $\bar{b} = 0.3$,  $\bar{d} = 1$, $\theta_0 = 0$, $\theta_\mathrm{ap} = \pi/12$ and inset the unit cell. }
\label{fig:ordercomparisonandapproxcomparison}
\end{figure}

 In this section we compare results from a full finite-element treatment for the original array problem \eqref{eq:nondimform} against results from the multipole-matched asymptotic system \eqref{eq:dispeqsystemgn}, as well as the leading order \eqref{eq:dispeqdipoleleading} and first-order \eqref{eq:dispeqfirstorder}  dispersion equation approximations. The objective is to examine the impact of varying    the central aperture angle $\theta_0$, the filling fraction $f$, and the half-angle aperture width $\theta_\mathrm{ap}$ on the first band and on the first band gap. We use finite-element results obtained from existing and readily available software     to independently benchmark the multipole-matched asymptotic results obtained here.
 
 In Figure \ref{fig:ordercomparisonandapproxcomparison}, we examine the first band surface of a representative resonator array possessing a moderate   half angle $\theta_\mathrm{ap} = \pi/12$,   apertures  located at $\theta_0 = 0$, and    filling fraction $f = \pi b^2/d^2 \approx 0.28 $: in Figure  \ref{fig:comparefigssubfig1} we compare results for the first band over the edge of the irreducible Brillouin zone (highlighted in Figure \ref{fig:schemsubfig2}) using both finite-element methods and our multipole-matched asymptotic system \eqref{eq:dispeqsystemgn}. Key features of the first band include   different low-frequency slopes along the high symmetry directions $\Gamma X$ and $\Gamma Y$, the existence of an almost flat band at the cutoff frequency along $MY$, and a saddle point frequency located at $X$. In this representative example, we find that the system \eqref{eq:dispeqsystemgn} is able to   describe the first band well over its entire frequency range (solid red line) within a dipole truncation, even with the use of lattice sum approximations (dashed black line), demonstrating that although the full system overestimates the frequency at $Y$, a dipole truncation gives a reasonable approximation over the entire Brillouin zone. Also superposed is the result within a quadrupolar truncation (solid green) which is an excellent approximation over the entire range, emphasising the importance of quadrupolar contributions to the modes as we approach the band edge. The adjacent Figure \ref{fig:comparefigssubfig2} overlays the finite-element result (blue line) with the lowest (solid red line) and first-order (dashed red line)  approximations for the dispersion equation. As described earlier, the lowest-order approximation is symmetric along all high symmetry directions (i.e., returns an isotropic approximation to the array), but is accurate only along $\Gamma Y$, being  unable to capture the saddle point at $X$; an improved description is obtained using the first-order approximation, with only a minor discrepancy in the band curvature along the $XM$ direction. In summary, the discrepancies in curvature along $XM$ are due to an absence of quadrupolar terms, the series expansions for the Bessel functions, and the lattice sum approximations, as shown in Figure  \ref{fig:comparefigssubfig1}.  
  
\begin{figure}[t]
\centering
\subfloat[Subfigure 1 list of figures text][]{
\includegraphics[width=0.475\textwidth]{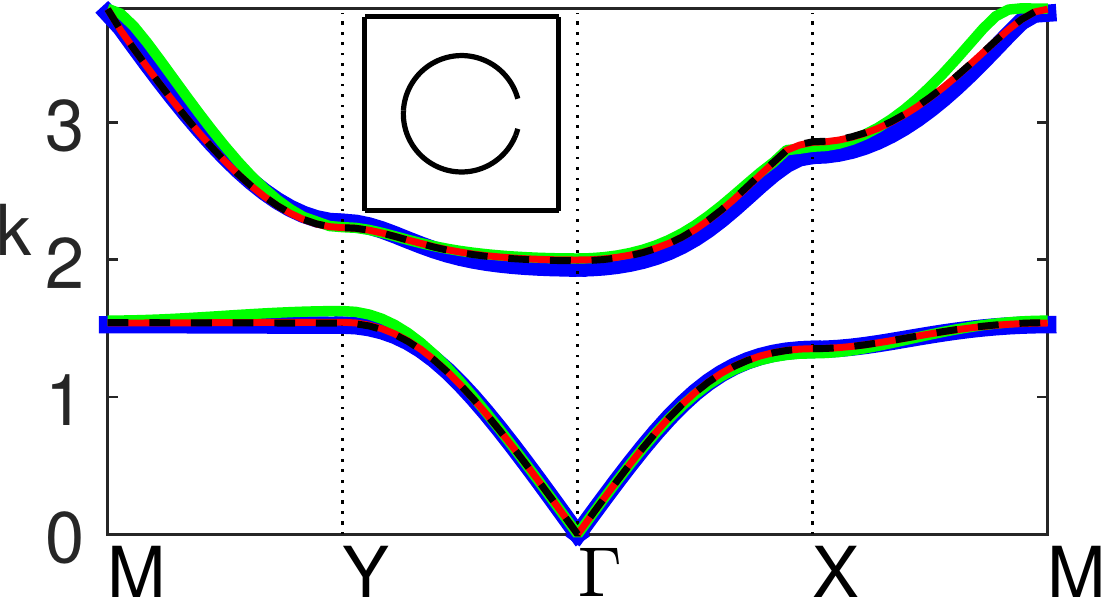}
\label{fig:varytheta0figssubfig1}}
\subfloat[Subfigure 2 list of figures text][]{
\includegraphics[width=0.475\textwidth]{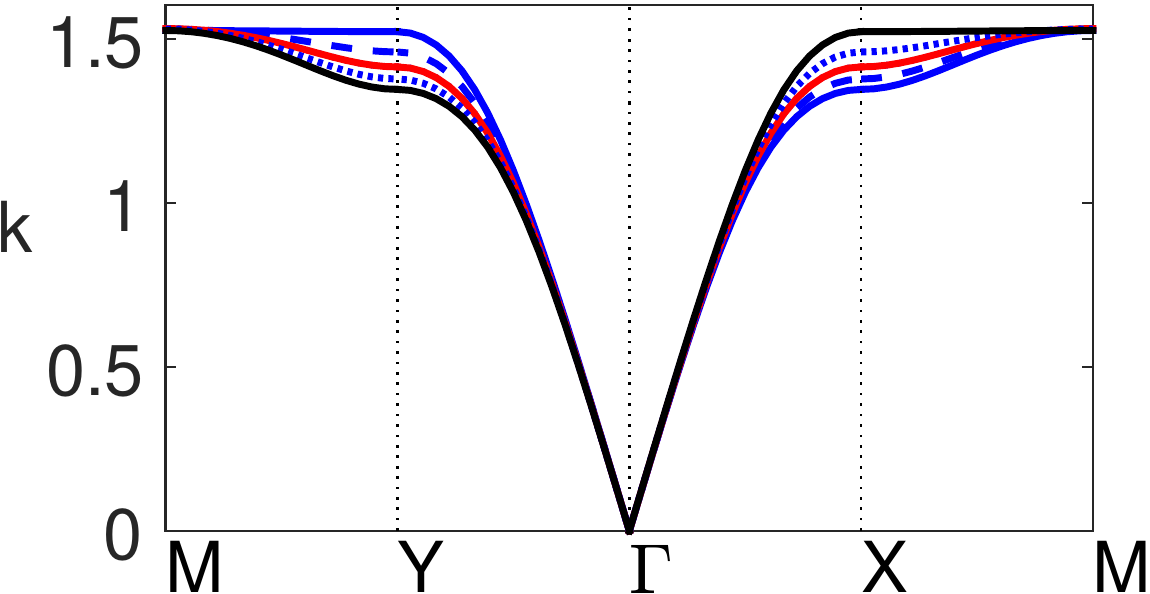}
\label{fig:varytheta0figssubfig2}} 
 
\caption{Band diagram  for   two-dimensional square array of thin-walled Helmholtz resonators as: \protect\subref{fig:varytheta0figssubfig1}
we increase the truncation of the multipole system \eqref{eq:dispeqsystemgn} from dipolar (green line), to quadrupolar (red line), and sextapolar (dashed black line), with finite-element result superposed (blue line) and fundamental unit cell inset; \protect\subref{fig:varytheta0figssubfig2} the central aperture angle is varied: $\theta_0 =0$ (blue line), $\theta_0 = \pi/6$ (dashed blue line), $\theta_0 = \pi/4$ (red line), $\theta_0 = \pi/3$ (dotted blue line), and $\theta_0 = \pi/2$ (black line) with curves   calculated using finite-element methods. In both figures we use $\bar{b} = 0.3$,  $\bar{d} = 1$, and $\theta_\mathrm{ap} = \pi/12$.    }
\label{fig:varytheta0figs}
\end{figure}

  In Figure \ref{fig:varytheta0figssubfig1} we compute the first two bands for the same resonator array configuration used in Figure \ref{fig:ordercomparisonandapproxcomparison}   using both finite-element and our multipole-matched asymptotic method    \eqref{eq:dispeqsystemgn}; we observe reasonable qualitative descriptions at dipolar truncation over both bands, with improvements in convergence over the first band along the $YM$ and $XM$ symmetry paths for quadrupolar truncations and higher. We observe that very good convergence for the (approximate) multipole-matched asymptotic treatment is achieved at quadrupolar truncation, and although the multipole-matched asymptotic system does not converge precisely to the finite element result, it still performs extremely well for an approximate description. Importantly, this figure suggests that the width of the first band gap may be determined with reasonable accuracy (with high enough truncation $L$), and that the greatest discrepancies in our model are observed on second band at the highest frequencies, as might be expected. In Figure \ref{fig:varytheta0figssubfig2}, we consider the effect of varying the central aperture angle $\theta_0$ on the band structure curvature (over the irreducible Brillouin zone for a high frequency configuration); results for several configurations in the range $0 \leq \theta_0 \leq \pi/2$ are superposed where we observe only   small changes in the curvature of the first band    for different $\theta_0$ angles.  Results from our multipole formulation match those obtained using  finite element methods, as expected, but are excluded here to avoid figure overcrowding. Accordingly, we consider $\theta_0 = 0$ in all other numerical results. For $\theta_0  = \pi/4$ we recover a symmetric band surface where the lowest order approximation \eqref{eq:dispeqdipoleleading}  possesses identical symmetry, however it overestimates   frequencies at $X$ and $M$; see Figure \ref{fig:comparefigssubfig2}. 
 
\begin{figure}[t]
\centering
\subfloat[Subfigure 1 list of figures text][]{
\includegraphics[width=0.475\textwidth]{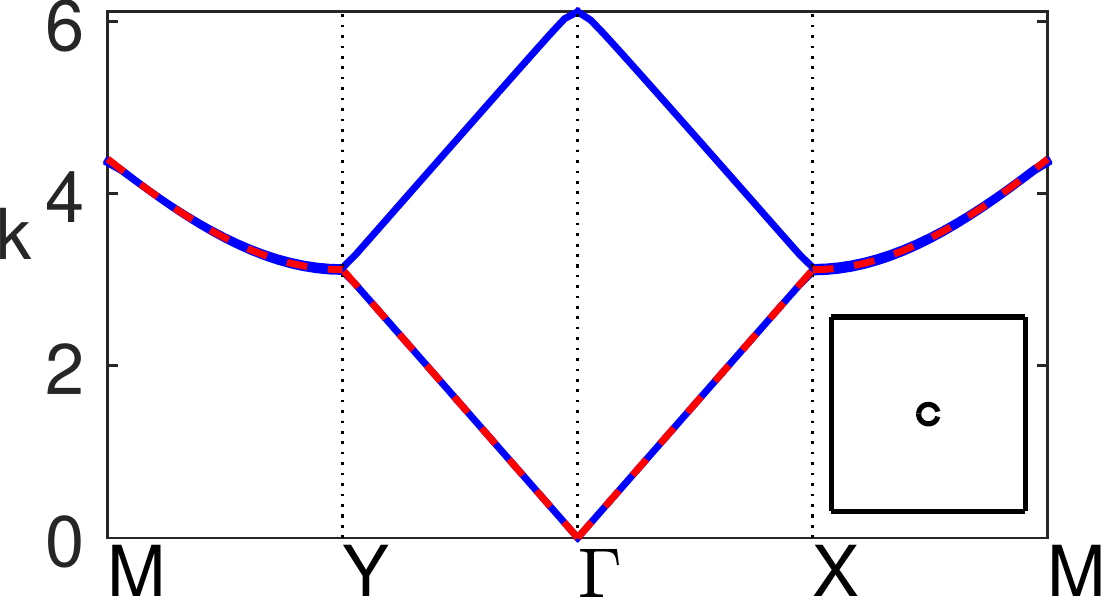}
\label{fig:ffracsubfig1}}
\subfloat[Subfigure 2 list of figures text][]{
\includegraphics[width=0.475\textwidth]{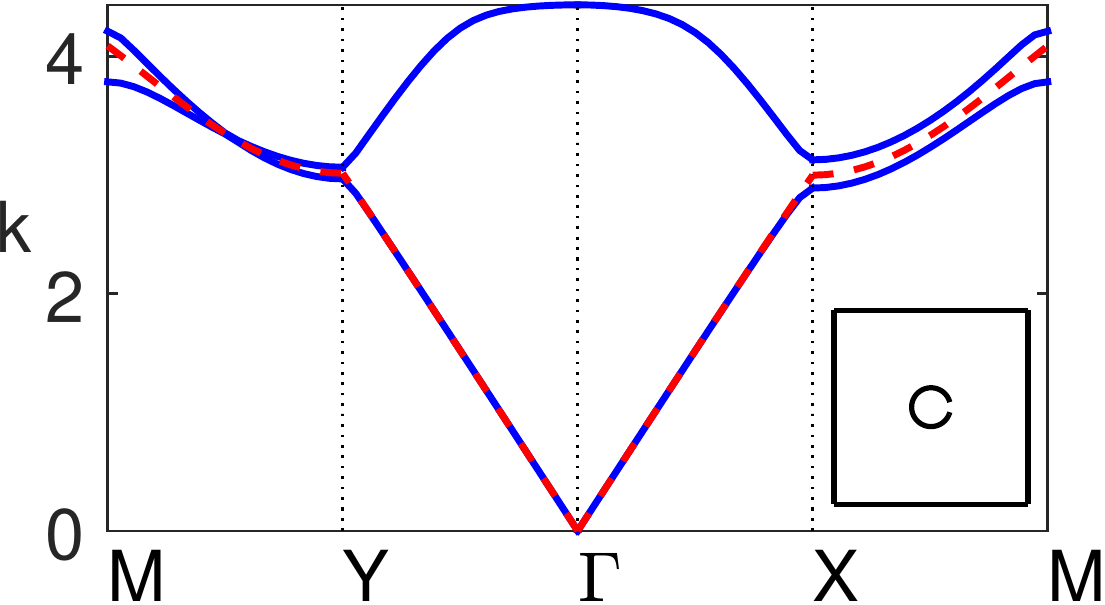}
\label{fig:ffracsubfig2}}\\
\subfloat[Subfigure 3 list of figures text][]{
\includegraphics[width=0.475\textwidth]{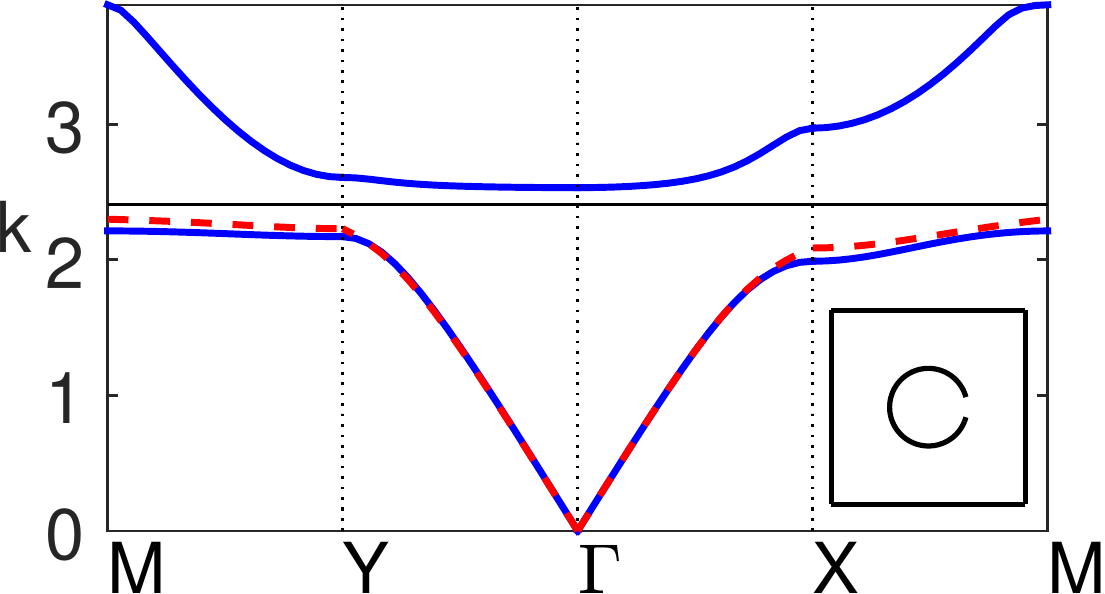}
\label{fig:ffracsubfig3}}
\subfloat[Subfigure 4 list of figures text][]{
\includegraphics[width=0.475\textwidth]{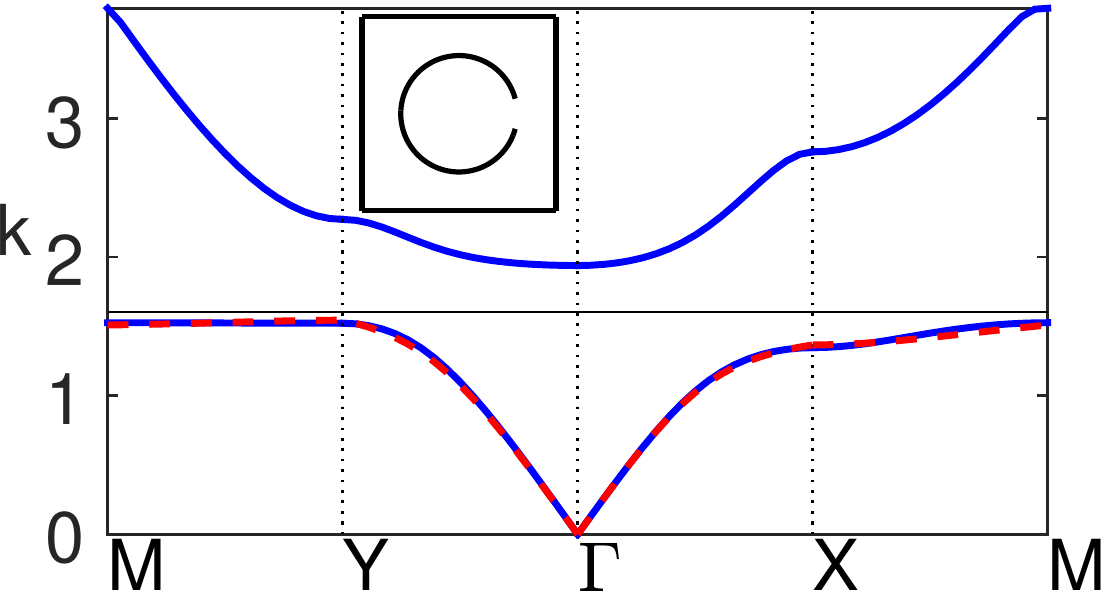}
\label{fig:ffracsubfig4}}\\
\subfloat[Subfigure 5 list of figures text][]{
\includegraphics[width=0.475\textwidth]{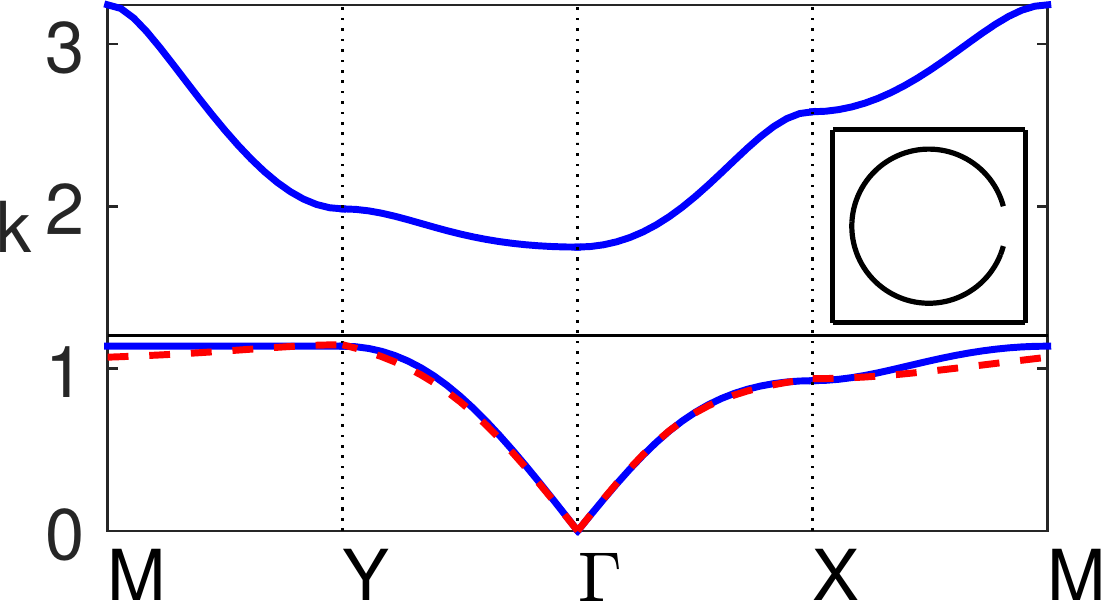}
\label{fig:ffracsubfig5}}
\subfloat[Subfigure 6 list of figures text][]{
\includegraphics[width=0.475\textwidth]{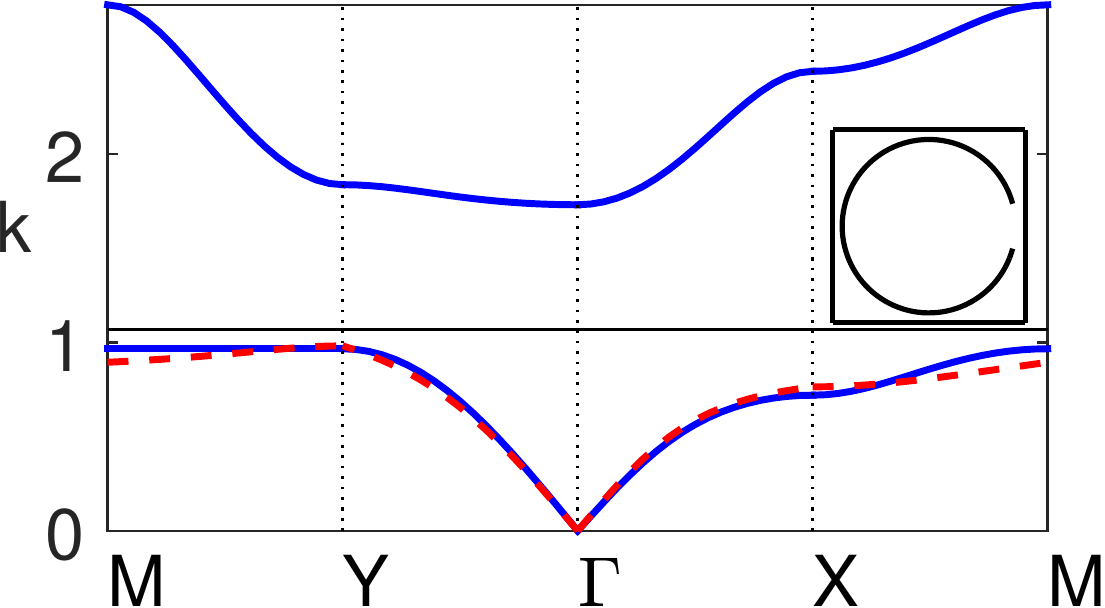}
\label{fig:ffracsubfig6}}
\caption{Band diagrams for a two-dimensional square array of thin-walled Helmholtz resonators  as the radius (i.e., filling fraction $f=\pi\bar{b}^2/\bar{d}^2$)  is varied: \protect\subref{fig:ffracsubfig1} $\bar{b} = 0.05$, \protect\subref{fig:ffracsubfig2}  $\bar{b} = 0.1$, \protect\subref{fig:ffracsubfig3}  $\bar{b} = 0.2$, \protect\subref{fig:ffracsubfig4}  $\bar{b} = 0.3$, \protect\subref{fig:ffracsubfig5}  $\bar{b} = 0.4$, \protect\subref{fig:ffracsubfig6}  $\bar{b} = 0.45$ with fundamental unit cells inset. Blue lines denote   from finite-element solution, red dashed lines denote results from the first-order correction \eqref{eq:dispeqfirstorder}, and black lines denote estimates for the edge of the band gap \eqref{eq:maxeval1stband}. In the above figures we   use $\bar{d} = 1$, $\theta_0 = 0$, and $\theta_\mathrm{ap} = \pi/12$.  }
\label{fig:ffracvariation}
\end{figure}

In Figure \ref{fig:ffracvariation} we examine   the performance of our first-order description   \eqref{eq:dispeqfirstorder} as the filling fraction is varied, for   the same configuration as in Figure \ref{fig:ordercomparisonandapproxcomparison} but as we vary the radius $\bar{b}$. We also superpose the estimate for the cutoff (band edge) frequency \eqref{eq:maxeval1stband} for instances where a band gap exists. We observe that the description for the first band works well both in the presence (here, $\bar{b}>0.1$) and absence (here, $\bar{b}<0.1$) of a band gap, although at higher filling fractions, the first-order description is unable to   capture the cutoff frequency  and the curvature with extreme precision, as we approach the $M$ point. In this figure we include   the first two bands to examine also  the effect of filling fraction on the first band gap; we find  that the gap width increases as $f$ increases, for fixed aperture width. Interestingly, the estimate for the cutoff frequency works best at moderate-to-high filling fractions (i.e., $f>0.28$), and that at very dilute filling fractions the bands approach the dispersion curves for free-space,    despite the presence of a resonator. Note that it is possible to determine an upper bound on the width of the first band gap by specifying the Bloch coordinate to lie at the $\Gamma$, $X$, $Y$, and $M$ points, solving for vanishing determinant,   choosing the second eigenvalue at these coordinates, and then selecting the minimum of these values. We advise   solving the full system \eqref{eq:dispeqsystemgn} to obtain these values and advise against the use of the dispersion equation \eqref{eq:dispeqfirstorder} for this purpose, as the accuracy of the second band values are not always assured within the   description.

\begin{figure}[t]
\centering
\subfloat[Subfigure 6 list of figures text][]{
\includegraphics[width=0.475\textwidth]{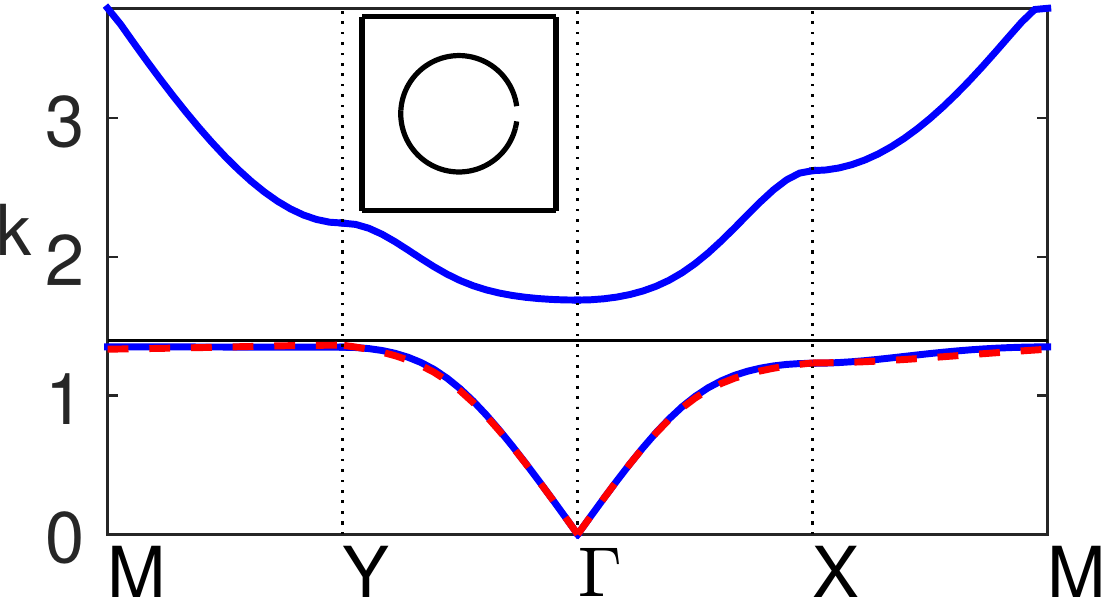}
\label{fig:thetaAsubfig6}}
\subfloat[Subfigure 1 list of figures text][]{
\includegraphics[width=0.475\textwidth]{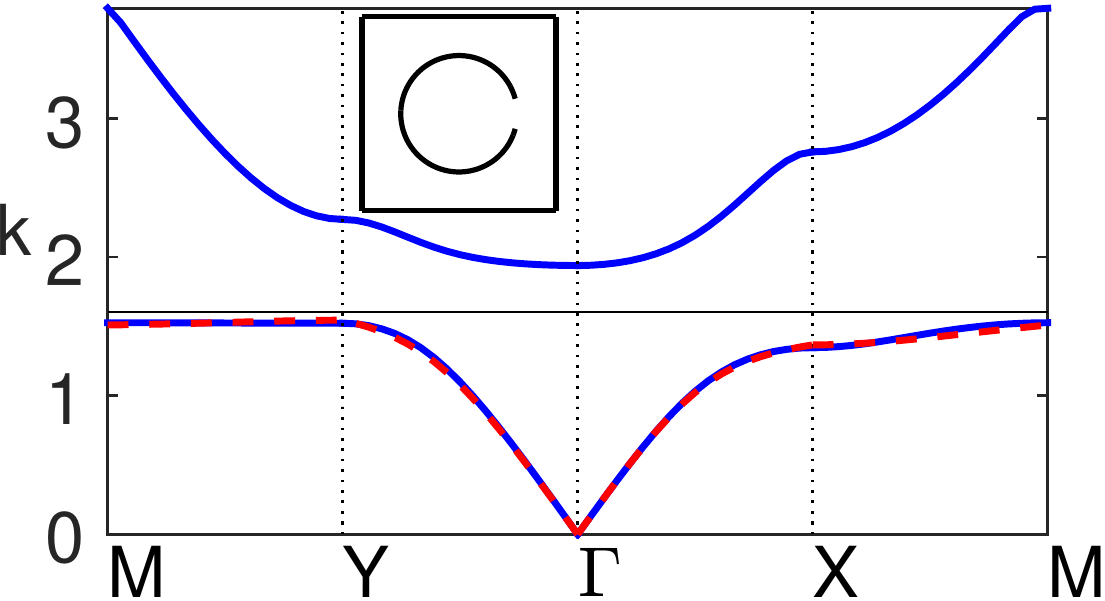}
\label{fig:thetaAsubfig1}}\\
\subfloat[Subfigure 2 list of figures text][]{
\includegraphics[width=0.475\textwidth]{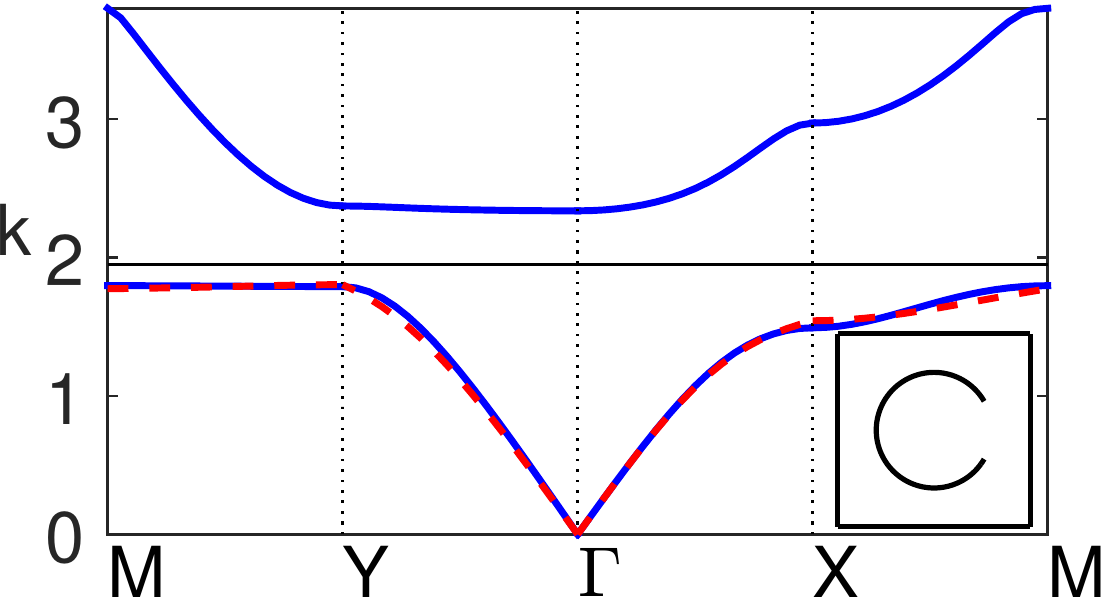}
\label{fig:thetaAsubfig2}}
\subfloat[Subfigure 3 list of figures text][]{
\includegraphics[width=0.475\textwidth]{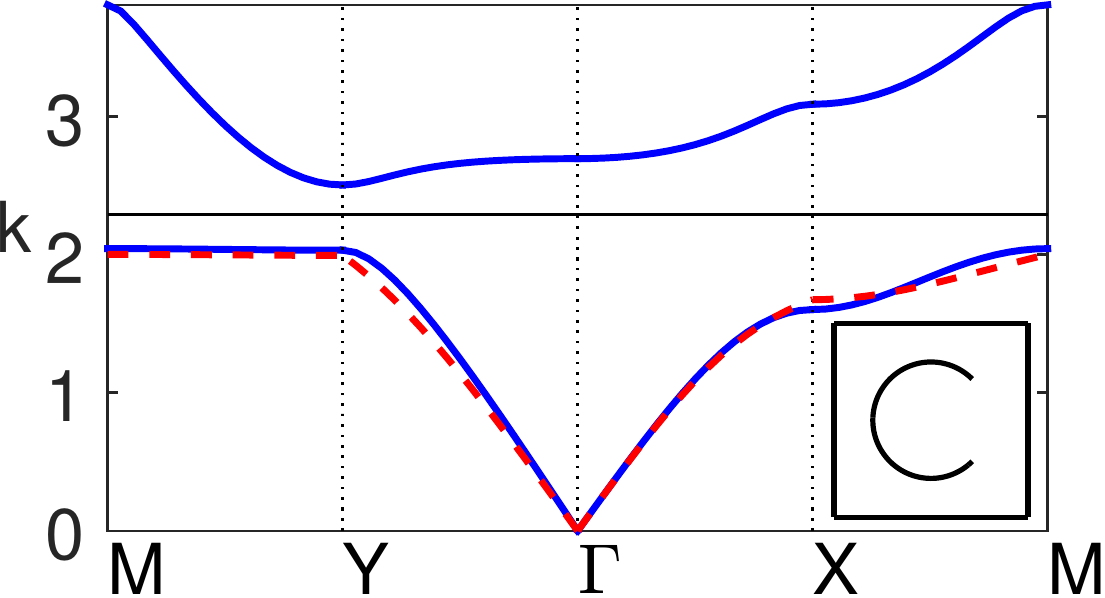}
\label{fig:thetaAsubfig3}}\\
\subfloat[Subfigure 4 list of figures text][]{
\includegraphics[width=0.475\textwidth]{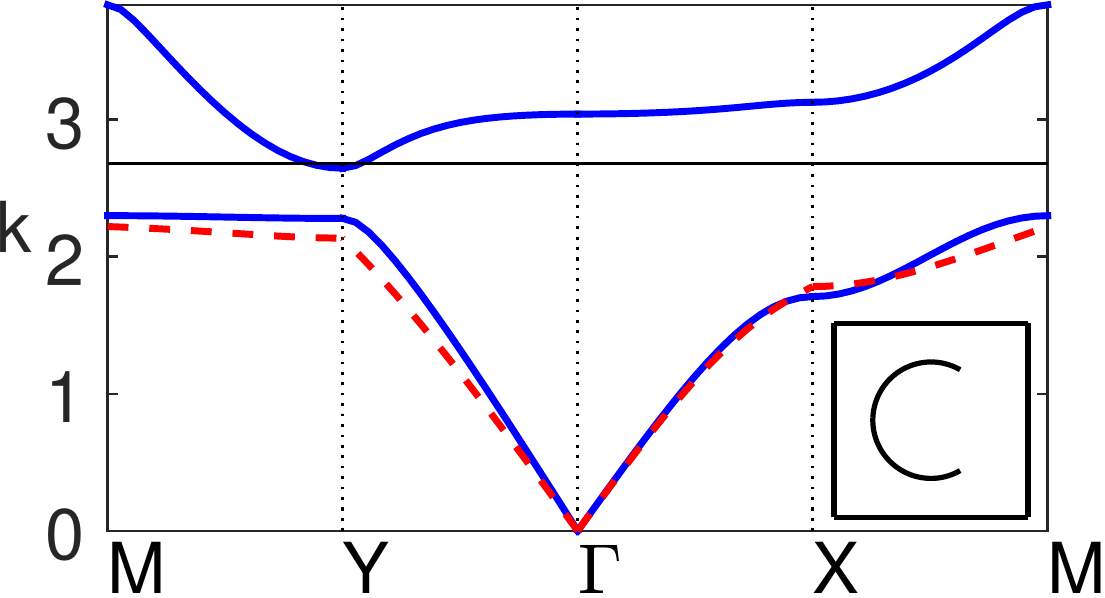}
\label{fig:thetaAsubfig4}}
\subfloat[Subfigure 5 list of figures text][]{
\includegraphics[width=0.475\textwidth]{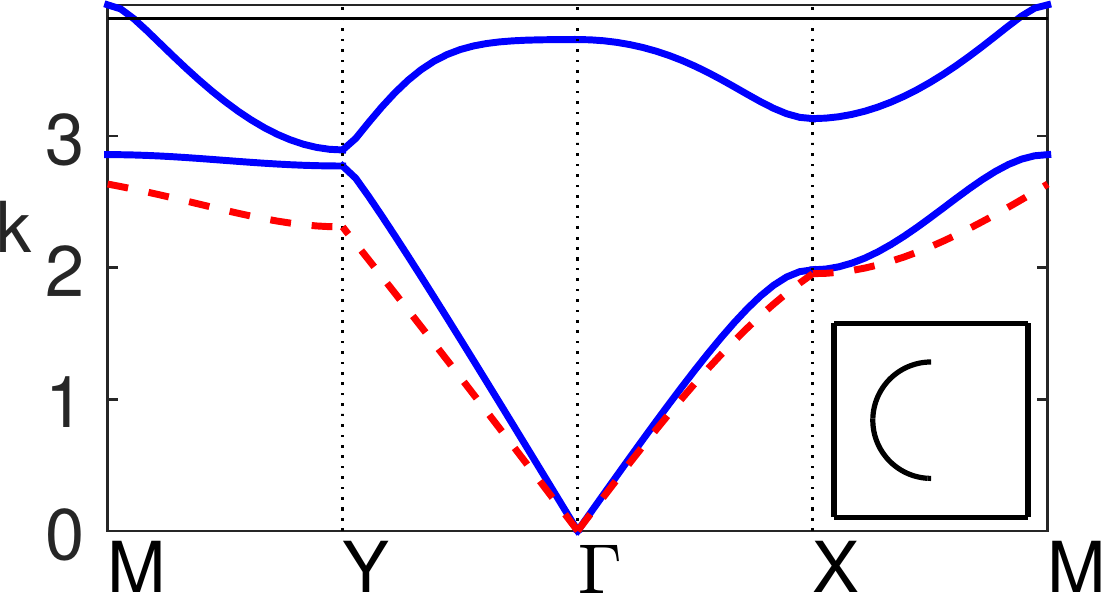}
\label{fig:thetaAsubfig5}}
\caption{Band diagrams for a two-dimensional square array of thin-walled Helmholtz resonators   as the aperture half-angle  is varied:  \protect\subref{fig:thetaAsubfig6}  $\theta_\mathrm{ap} = \pi/24$, \protect\subref{fig:thetaAsubfig1} $\theta_\mathrm{ap} = \pi/12$, \protect\subref{fig:thetaAsubfig2} $\theta_\mathrm{ap} = \pi/6$, \protect\subref{fig:thetaAsubfig3}  $\theta_\mathrm{ap} = \pi/4$, \protect\subref{fig:thetaAsubfig4}  $\theta_\mathrm{ap} = \pi/3$, \protect\subref{fig:thetaAsubfig5}  $\theta_\mathrm{ap} = \pi/2$ with fundamental unit cells inset. Blue lines denote results from finite-element methods, red dashed lines denote results from the first-order correction \eqref{eq:dispeqfirstorder}, and black lines denote estimates for the edge of the band gap \eqref{eq:maxeval1stband}. In the above figures we use $\bar{d} = 1$, $\theta_0 = 0$, and $\bar{b} = 0.3$.   }
\label{fig:thetaAvariation}
\end{figure}

 In Figure \ref{fig:thetaAvariation} we investigate how well the first-order description \eqref{eq:dispeqfirstorder} works with increasing aperture size, that is, we examine   the same configuration as in Figure \ref{fig:ordercomparisonandapproxcomparison} but now vary the  half-angle $\theta_\mathrm{ap}$. We find that   our description works well up to half-angles of $\theta_\mathrm{ap} \approx \pi/4$, which is perhaps remarkable  for a system formally derived in the vanishing aperture limit. We observe that the first-order description breaks down    when a clear minimum of the second band surface appears at the $Y$ high-symmetry coordinate, rather than   at  the $\Gamma$ point. It also corresponds with the estimated band maximum appearing at approximately the midpoint of the band gap, which closes with increasing aperture size. Finally, we remark that our description still holds moderately well  up to a larger half-angle of $\theta_\mathrm{ap} = \pi/3$,   along the $\Gamma X$ direction.
 
%
%


Having examined the parameter ranges over which our system and dispersion equation are valid, we now investigate the effects of wall thickness on results for Helmholtz resonator arrays.

 \section{Extension to thick-walled resonators}
 \label{chap:thicksec}

 We now pose  the governing equations for the thick-walled resonator problem  shown in Fig.~\ref{fig:schemsubfig1thick}, in terms of the non-dimensional coordinates   \eqref{eq:nondimk}. By thick-walled, we mean a cylinder whose aperture arc length, $2\overline{\ell}$, is of the same order as its thickness. This has an identical structure to   \eqref{eq:nondimform} earlier but now possesses  a modified Neumann boundary condition in  the form
\begin{subequations}
\label{eq:nondimformthick}
\begin{align}
\label{eq:neumannthick}
\frac{\partial\phi}{\partial r}  \bigg|_{S_\mathrm{T}} &= 0, 
\end{align}
\end{subequations}
where    $S_\mathrm{T}$ denotes the thick-walled Helmholtz resonator. 
The definition is chosen to  ensure  that the resonator walls in the neck are parallel to one another, as shown in Fig.~\ref{fig:schemsubfig1thick}, and admits the inner problem domain presented in Fig.~\ref{fig:schemsubfig2thick}.  We write the non-dimensional inner radius $a= b-2 h \varepsilon$, where $\varepsilon = \ell$ is  the aperture arc half-length at the outer radius $b$, and $h$ is  the aspect ratio of the channel (resonator neck). Note also that the definition of the inner radius given above implicitly treats the aperture neck length $(b-a)$ as $O(\varepsilon)$, which  later ensures that the filling fraction is held constant (see below).

\subsection{Inner problem formulation}
\label{chap:innerthick}
\begin{figure}[t]
	\centering
	 \subfloat[Subfigure 1 list of figures text][]{
\includegraphics[width=0.475\textwidth]{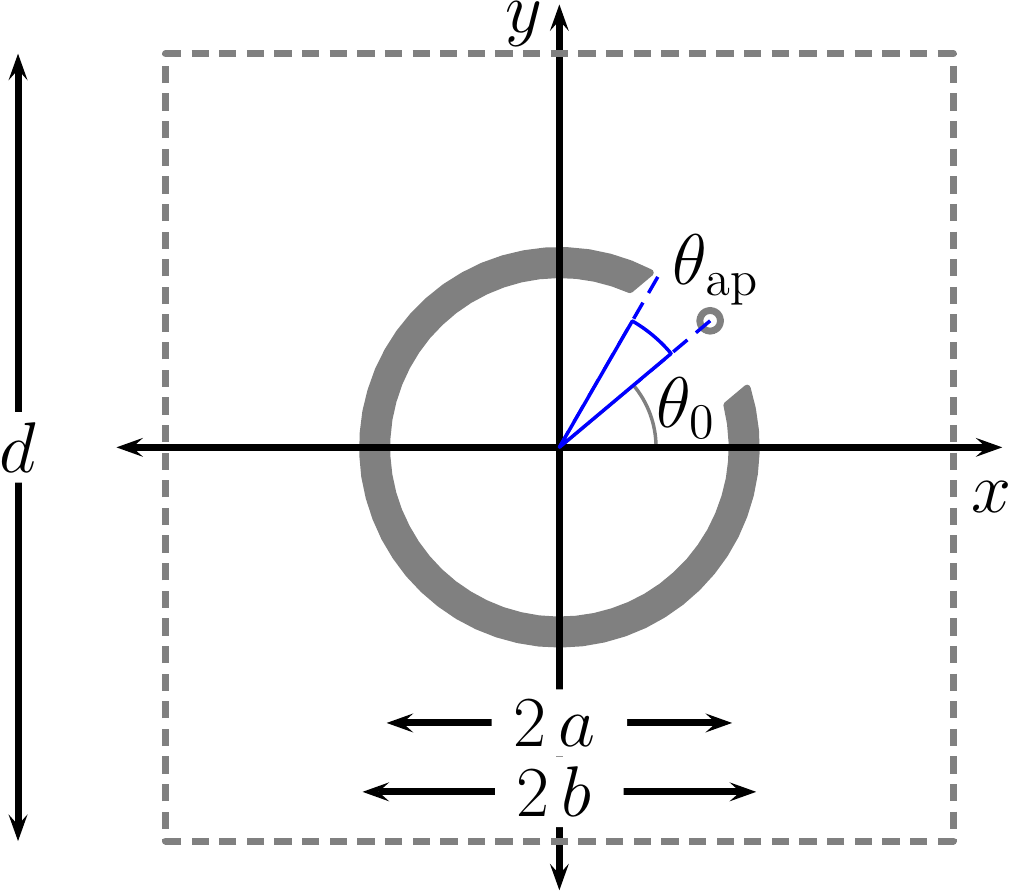}
\label{fig:schemsubfig1thick}}\hspace{10mm} 
\subfloat[Subfigure 2 list of figures text][]{
\includegraphics[width=0.41\textwidth]{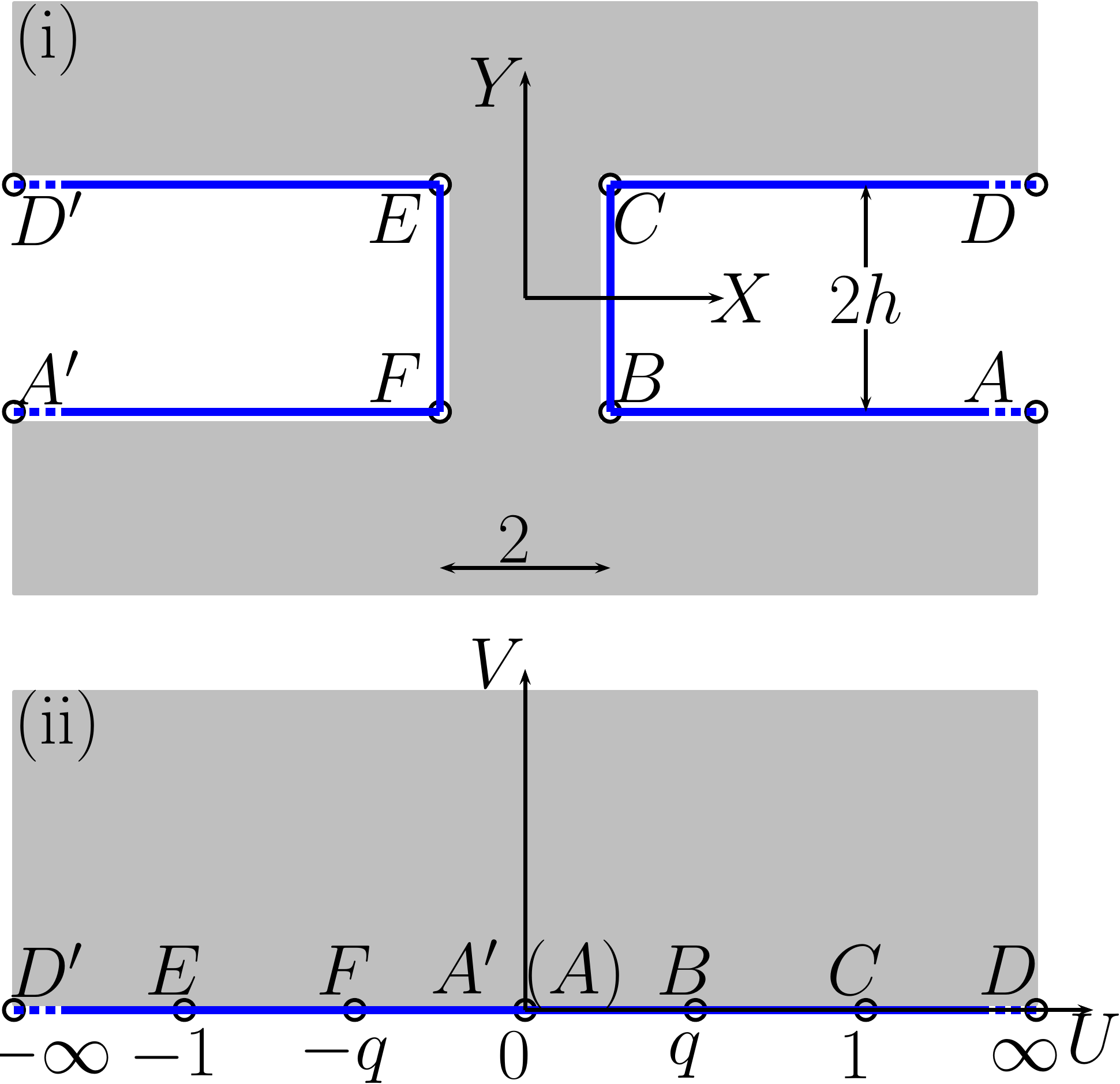}
\label{fig:schemsubfig2thick}}

	\caption{(a) Fundamental unit cell for a square array of period $d$ containing a thick-walled cylindrical resonator with inner radius $a$, outer radius $b$, and an aperture centred at $\theta_0$ with  half-angle   $\theta_\mathrm{ap} = \ell/b$, where the outer aperture arc length is  $2\ell$; (b)(i) Inner problem geometry with  unbounded polygonal (fluid) domain   overlaid in gray; (b)(ii) Inner problem geometry obtained  via the Schwarz--Christoffel mapping \eqref{eq:schwarz-christ-thick}; the capital letters $A,\ldots,D$ and $A^\prime,\ldots,D^\prime$ denote the points of correspondence in the $Z (=X+\rmi Y)$ and $W (=U+\rmi V)$ complex planes. \label{fig:resonatorarraythick}}
\end{figure}

As before, we first rotate and translate the array via the operation $(\tilde{x},\tilde{y})\mapsto (x\sin\theta_0 - y\cos\theta_0,x\cos\theta_0 + y\sin\theta_0 - b + \ell h)$ so that the origin in tilde coordinates is centred and oriented on the aperture in the fundamental cell. As in Section \ref{chap:innerthin}, we   introduce the inner  scaling \eqref{eq:innerexp}  and  a regular expansion for $\phi$  to  obtain the leading-order system    
\begin{subequations}
\begin{align}
(\partial_X^2 + \partial_Y^2) \Phi = 0, \quad \mbox{ for } X  \in \mathbb{R}^2 \backslash S^\mathrm{in}_\rmT, \\
\partial_{N} \Phi   = 0, \quad \mbox{ for } X  \in S^\mathrm{in}_\rmT,
\end{align}
\end{subequations}
 where  $\partial_{N}$ denotes the   normal derivative,   the resonator walls are defined by \newline $S^\mathrm{in}_\rmT = \left\{(X ,Y ): |X|\geq 1 \times Y \in \left[-h,h\right]    \right\}$ as shown in Fig.~\ref{fig:schemsubfig2thick}, and we   omit  the subscript for $\Phi_0$. Next we introduce the  Schwarz--Christoffel mapping \cite{fuchs1964functions} between the $Z$ and $W$ planes:
 \begin{subequations}
 \begin{equation}
 \label{eq:schwarz-christ-thick}
Z(W) =\frac{  \int_{1}^{W}\dfrac{\sqrt{(\zeta^2-1)(\zeta^2-q^2)}}{\zeta^2}  \, \rmd \zeta+ \int_{-q}^{W} \dfrac{\sqrt{(\zeta^2-1)(\zeta^2-q^2)}}{\zeta^2} \, \rmd \zeta   }{  \int_{-q}^{q} \, \dfrac{\sqrt{(\zeta^2-1)(\zeta^2-q^2)}}{\zeta^2}  \, \rmd \zeta},
\end{equation}
where the parameter $q$ is related exactly to the specified aspect ratio $h$ via
\begin{equation}
\label{eq:scmapqh}
h = \frac{1}{2}\left[ 2 E(q^2) + (q^2-1) K(q^2)\right]^{-1}\left[- 2E(1-q^2) + (1+q^2) K(1-q^2)  \right],
\end{equation}
\end{subequations}
 and $E(k)$ and $K(k)$ are complete Elliptic integrals of the first and second kind, respectively. Note that the aspect ratio $h$  cannot be too large  as $q$ vanishes exponentially in the limit of large $h$ (for example, for $h=4$ we have   $q\approx 1.8879\times10^{-6}$) which may cause accuracy issues from a numerical perspective.
    Hence, the treatment we outline  here implicitly  assumes thick-walled resonators where the channel aspect ratio $h$ is not too large (in fact, we may consider it to be $O(1)$).

Subsequently, using \eqref{eq:schwarz-christ-thick} we map the problem of solving Laplace's equation in the physical junction  domain $\mathbb{R}_2\backslash S_\rmT^\mathrm{in}$ shown  in Figure \ref{fig:schemsubfig2thick} to solving Laplace's equation in the upper-half plane of the $W$-plane shown in Figure \ref{fig:schemsubfig2thick}(ii), where a vanishing Neumann condition is imposed along the real line.  For the latter problem we may immediately offer a solution in the form
\begin{equation}
\Phi(W) = C_3 \mathrm{Re}\left\{ \log W \right\} + C_4,
\end{equation}
where from the leading order asymptotic form   for the mapping \eqref{eq:schwarz-christ-thick} above
\begin{equation}
\lim_{W \rightarrow 0} Z(W) \sim  \frac{\mathcal{C} (q) q}{W}, \quad \mbox{ and } \quad \lim_{W \rightarrow \infty} Z(W) \sim \mathcal{C}(q) W,
\end{equation}
we obtain the leading-order   result  in the original inner region  as
\begin{equation}
\label{eq:rhsasythick}
\left( \lim_{R\rightarrow \infty}\Phi \right) \bigg|_{R = \tilde{r}/\varepsilon}\sim 
\begin{cases}
C_3 \mathrm{Re}\left[ \log( \tilde{r} ) - \log\left\{\mathcal{C} (q) \varepsilon\right\} \right] + C_4,   & Z \in \mathbb{C}^\rmU, \vspace{1mm} \\
C_3 \mathrm{Re}\left[   \log\left\{q \,\mathcal{C} (q) \varepsilon\right\} -  \log(  \tilde{r} )  \right] + C_4,  & Z \in \mathbb{C}^\rmL,
\end{cases}
\end{equation}
 where 
\begin{equation}
\mathcal{C}(q) =  ( 2 E(q^2) + (q^2-1)K(q^2) )^{-1},
\end{equation}
and we reintroduce tilde notation as before. Note that when $q=1$ we have $\mathcal{C}  = 1/2$ and $h=0$ to recover the asymptotic form for the thin-walled resonator outlined before.  Next we outline modifications to  the outer problem, specifically, the outer interior problem solution.

\subsection{Outer interior problem formulation}
 The derivation of the outer interior solution proceeds analogously to that given in Section \ref{chap:outthin}\ref{sec:outintthin}, but with the replacement coordinates and parameters   $\tilde{r} \mapsto \check{r}$, $b \mapsto a$, and $Q_m \mapsto \check{Q}_m$, where we define
 \begin{equation}
  \check{Q}_m =  	\rmJ_m(a)\rmH_m^{(1)\prime}(a) + \rmJ_m^\prime(a) \rmH_m^{(1)}(a),
\end{equation}
to   obtain
\begin{equation}
\label{eq:genformneumannsatBthick}
\phi_\mathrm{int} = B \, \rmH_0^{(1)}(\check{r}) - \frac{B}{2}\sum_{n=-\infty}^{\infty} \frac{\check{Q}_n}{\rmJ_n^\prime(a)} \rmJ_n(r) \rme^{\rmi n (\theta-\theta_0)}.
\end{equation}
Hence, as we approach the resonator neck from the interior and exterior domains, the outer solution now takes the form
\begin{equation}
\label{eq:outerasympsbothdirthick}
\lim_{\theta \rightarrow \theta_0} \lim_{r \rightarrow b,a}  \phi_\mathrm{out}  \sim
\begin{cases}
 \dfrac{2\rmi A}{\pi} \left[ \gamma_\rme - \dfrac{\rmi \pi}{2} + \log\left(\dfrac{\tilde{r}}{2}\right) \right]  + \sum\limits_{n=-\infty}^{\infty}b_n    \rmY_n(b)     \rme^{\rmi n \theta_0}
  \\ 
 \hspace{40mm}- 
\sum\limits_{n=-\infty}^{\infty} \left\{     \dfrac{A Q_n}{2  }    + b_n \rmY_n^\prime(b)    \rme^{ \rmi n    \theta_0 } \right\}   \dfrac{\rmJ_n(b)}{\rmJ_n^\prime(b)}  
 , &  r \downarrow b, \\
\dfrac{2\rmi B}{\pi} \left[ \gamma_\rme - \dfrac{\rmi \pi}{2} + \log\left(\dfrac{\check{r}}{2}\right) \right] - \dfrac{B}{2}\sum\limits_{n=-\infty}^{\infty} \dfrac{\check{Q}_n}{\rmJ_n^\prime(a)} \rmJ_n(a), \phantom{\bigg|^b} 	& r \uparrow a,
\end{cases}
\end{equation}
which is the analogue to the   thin-walled expression in \eqref{eq:outerasympsbothdir} given earlier, but with the addition of the inner wall radius $a=b-2h\varepsilon$.

\subsection{Matching procedure for thick-walled resonators}
Having obtained  the    inner and outer asymptotic representations   \eqref{eq:rhsasythick} and \eqref{eq:outerasympsbothdirthick}, we now match inner fields in the upper- and lower-half planes to outer fields as  $r \downarrow b$ and $r \uparrow a$, respectively. As before, after matching logarithmic and non-logarithmic terms we obtain a system of equations, from which we find that $B=-A$ once more, but obtain an updated relationship between $A$ and $b_n$ analogous to that given in   \eqref{eq:Abnrel} but  with the replacement $h_\varepsilon \mapsto \check{h}_\varepsilon$
where
\begin{equation}
\label{eq:hepsfullthick}
\check{h}_\varepsilon =\dfrac{4\rmi  }{\pi} \left[ \gamma_\rme - \dfrac{\rmi \pi}{2} + \log\left(\dfrac{\varepsilon \, \mathcal{C}(q) \sqrt{q}}{2}\right) \right] - \frac{1}{2} \sum\limits_{n=-\infty}^{\infty} \dfrac{Q_n  \rmJ_n(b)}{\rmJ_n^\prime(b)}  
- \frac{1}{2} \sum\limits_{n=-\infty}^{\infty} \dfrac{\check{Q}_n  \rmJ_n(a)}{\rmJ_n^\prime(a)}.
\end{equation} 
Thus, we obtain an eigenvalue problem for  the thick-walled resonator case that is identical to   \eqref{eq:dispeqsystemgn}, but with the simple   replacement  $h_\varepsilon \mapsto \check{h}_\varepsilon$. This highlights a significant advantage of the present approach, as all local details of the neck geometry are contained in the single term $h_\varepsilon$.

\subsection{Leading and first-order systems}
The asymptotic form for $\check{h}_\varepsilon$ both within a dipolar truncation and in the vanishing $b$ limit,  can easily be shown to take the form
\begin{equation}
\lim_{b\rightarrow 0}\check{h}_\varepsilon \approx   \frac{4\rmi}{\pi}\left[ \log\left(\frac{\varepsilon  \sqrt{q} \, \mathcal{C}(q)}{\sqrt{ab}} \right) + \frac{1}{2}\left( \frac{1}{a^2} + \frac{1}{b^2}\right) - \frac{1}{8} \right].
\end{equation}
 Therefore after introducing   $\check{h}_\varepsilon = 4\rmi \check{f}_\varepsilon/(\pi b^2)$ as before we obtain
 \begin{equation}
 \label{eq:fepsasymptthick}
\check{f}_\varepsilon =  \left[  \frac{1}{2}\left( 1 + \frac{b^2}{a^2}\right)    -\frac{b^2}{8} + b^2\log \left( \frac{\varepsilon \sqrt{q} \, \mathcal{C}(q)}{\sqrt{ab}} \right)\right] \sim  \frac{1}{2}\left( 1 + \frac{b^2}{a^2}\right)     + b^2\log \left( \frac{\theta_\mathrm{ap} \sqrt{b} \sqrt{q} \, \mathcal{C}(q)}{\sqrt{a }} \right)  ,
\end{equation}
which is the analogue to the earlier thin-walled expression \eqref{eq:fepsasympt}. Hence we obtain the same dispersion equations as before, i.e.,  the leading-order expression in \eqref{eq:dispeqdipoleleading} and the   first-order correction expression in \eqref{eq:dispeqfirstorder}, but with the replacement $f_\varepsilon \mapsto \check{f}_\varepsilon$, that is, for thick-walled resonators the first-order   dispersion equation is 
\begin{equation}
\label{eq:dispeqfirstorderthick}
k_\rmB^2 = \frac{(f+1) \left[b^2 f  (2 f-1) +   \check{f}_\varepsilon   (f+1) (2 \check{f}_\varepsilon f-2 \check{f}_\varepsilon-2 f+1)\right]}{b^2 f  \cos (2 \left[\theta_0-\theta_\rmB\right])+b^2 f^2+  \check{f}_\varepsilon (2 \check{f}_\varepsilon-1) \left(f^2-1\right)}.
\end{equation}
For reference, the analogue to the lowest-order approximation for thin resonators \eqref{eq:correctedsls}    follows straightforwardly, and 
   finally,  the dipolar estimate for the cut-off frequency of the first band now takes the form  
\begin{equation}
\label{eq:maxeval1stbandthick}
  k_\mathrm{max} = \frac{2}{\bar{a}}\sqrt{\frac{1}{1-8\log ( \theta_\mathrm{ap} \sqrt{\bar{b}} \sqrt{q} \, \mathcal{C}(q)/\sqrt{\bar{a} } )}}.
\end{equation}

%

\subsection{Numerical results}
In this section we  briefly examine the validity of the multipole-matched asymptotic eigenvalue problem \eqref{eq:dispeqsystemgn} with the update $h_\varepsilon \mapsto \check{h}_\varepsilon$ described  in \eqref{eq:hepsfullthick}, as well as the new first-order approximation for the first band in \eqref{eq:dispeqfirstorderthick}.

\begin{figure}[t]
\centering
\subfloat[Subfigure 6 list of figures text][]{
\includegraphics[width=0.475\textwidth]{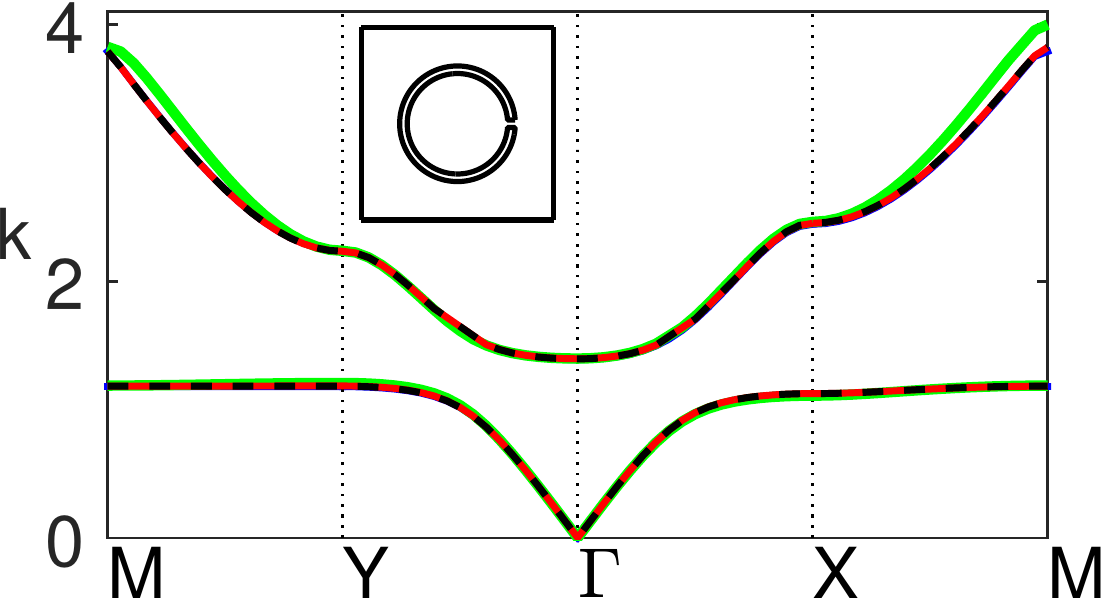}
\label{fig:thickmultipolesubfig1}}
\subfloat[Subfigure 1 list of figures text][]{
\includegraphics[width=0.475\textwidth]{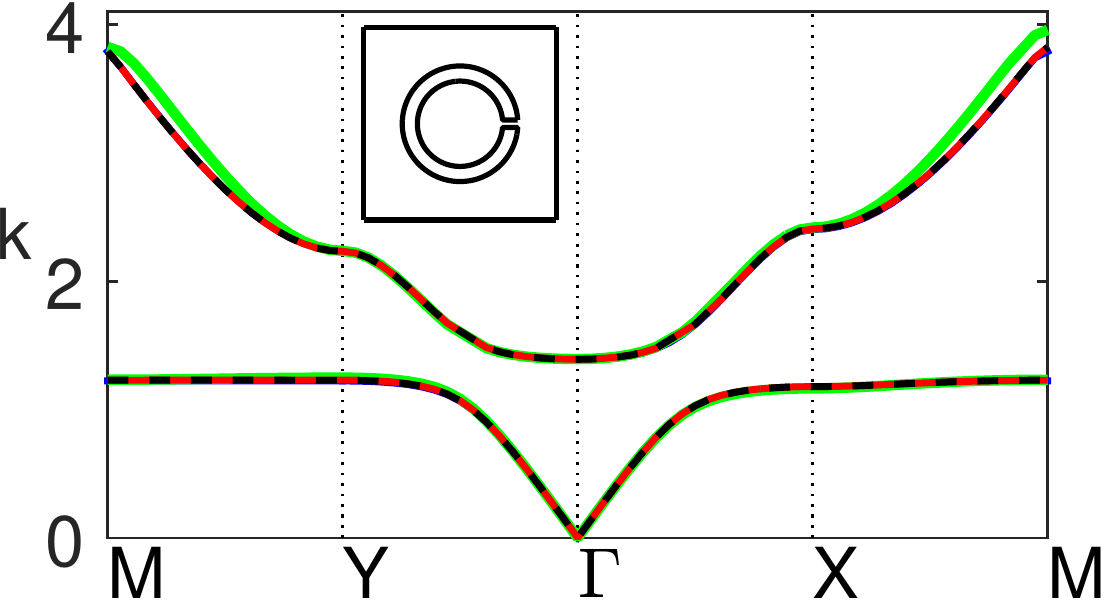}
\label{fig:thickmultipolesubfig2}}\\
\subfloat[Subfigure 2 list of figures text][]{
\includegraphics[width=0.475\textwidth]{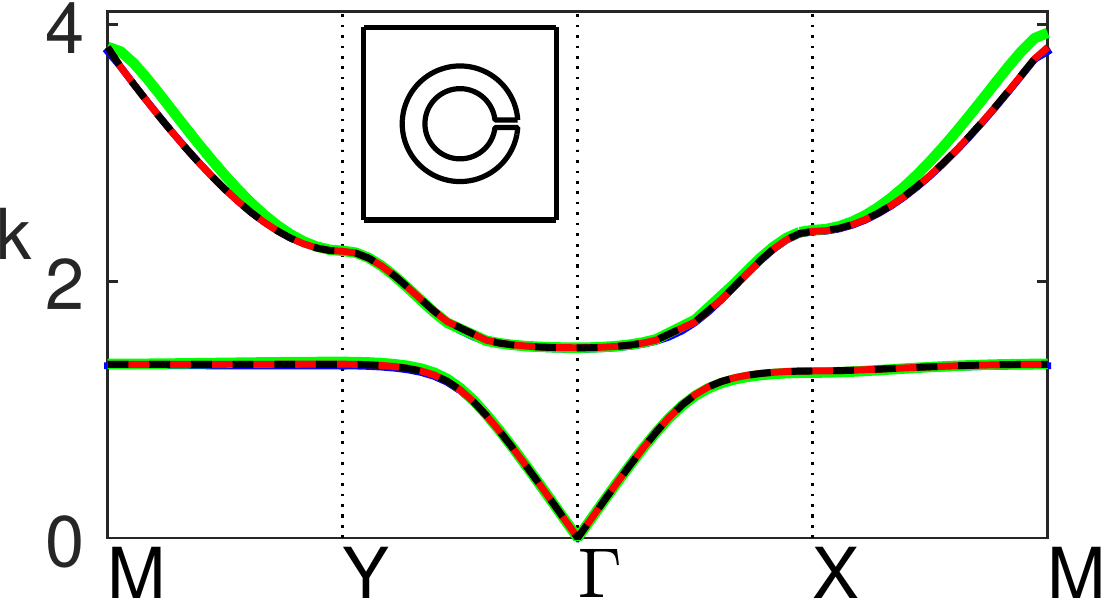}
\label{fig:thickmultipolesubfig3}}
\subfloat[Subfigure 3 list of figures text][]{
\includegraphics[width=0.475\textwidth]{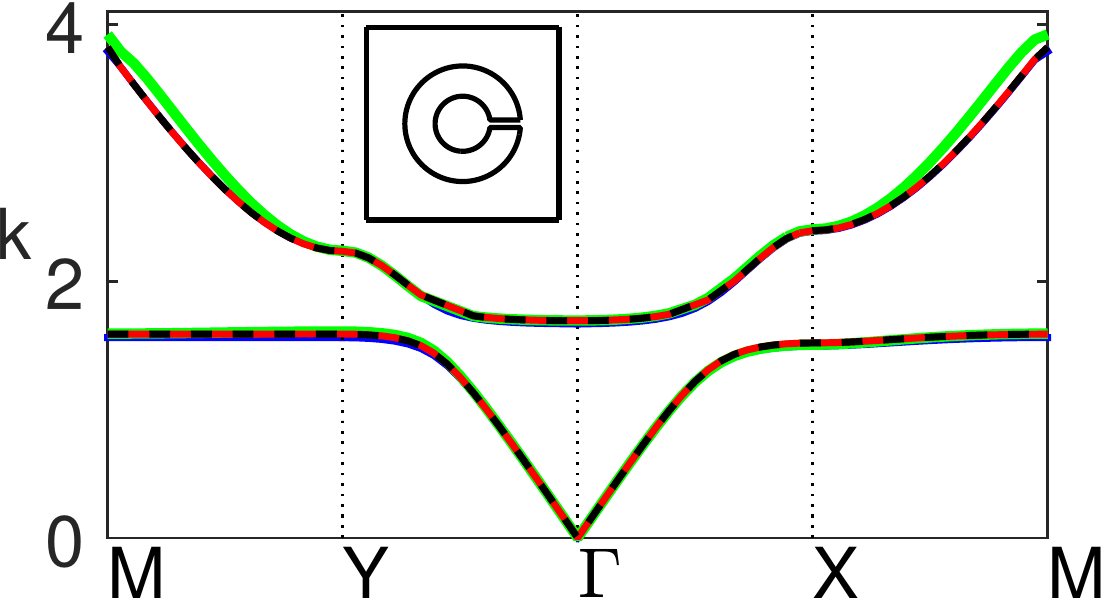}
\label{fig:thickmultipolesubfig4}}
\caption{Band diagrams for a two-dimensional square array of thick-walled Helmholtz resonators  as  the relative wall thickness (or aspect ratio) $h$   is increased:  \protect\subref{fig:thickmultipolesubfig1}  $h=1$, \protect\subref{fig:thickmultipolesubfig2} $h=2$, \protect\subref{fig:thickmultipolesubfig3} $h=3$, and \protect\subref{fig:thickmultipolesubfig4} $h=4$, with fundamental unit cells inset. Multipole results for dipole (green line), quadrupole (red line), and sextapole (dashed black line) truncations are   superposed, in addition to finite-element results (blue line).  In the above figures we use $\bar{d} = 1$, $\theta_0 = 0$,   $\bar{b} = 0.3$, and $\theta_\mathrm{ap} = \pi/48$.}
\label{fig:thickness48hvariation}
\end{figure}

In Figure \ref{fig:thickness48hvariation}, we compare results from our eigenvalue formulation  \eqref{eq:dispeqsystemgn} for various truncations (dipole $L=1$, quadrupole $L=3$, and sextapole $L=5$) against results obtained using finite-element methods, as we vary the thickness, or equivalently, the width of the resonator neck. We observe that  for this narrow half-angle $\theta_\mathrm{ap} = \pi/48$, we achieve excellent agreement with finite-element benchmark results and rapid convergence, with results indistinguishable above dipole truncation and higher, for both bands. We discuss  the band gap evolution, with increasing $h$, below in Figure \ref{fig:compallfigs}. In Figure \ref{fig:thickness24hvariation} we examine the efficacy of  the first-order approximation \eqref{eq:hepsfullthick} for a slightly wider half-angle $\theta_\mathrm{ap} = \pi/24$, and observe very good agreement over a   range of thickness $h$ values; it becomes clear by Figure \ref{fig:thickmultipole24subfig4} that for very large $h$ the model is no longer able to accurately describe the first band towards the band edge, but that at longer wavelengths, the description is still accurate. In Figure \ref{fig:compallfigs} we superpose the bands from Figures  \ref{fig:thickness48hvariation} and   \ref{fig:thickness24hvariation} to describe the influence of increasing thickness, and find that it acts to close the band gap, to steepen the slope of the first band at lower frequencies, and to translate the frequency range of the gap. This result is entirely consistent with the idea that as the thickness increases, the interior resonator shrinks so that the cut-off frequency increases. As before, we see the band gap closing at the $Y$ high-symmetry point, emphasising as before the important point that the upper bound of the band gap cannot always be assessed from examining the spectrum at the $\Gamma$ point alone.

\begin{figure}[t]
\centering
\subfloat[Subfigure 6 list of figures text][]{
\includegraphics[width=0.475\textwidth]{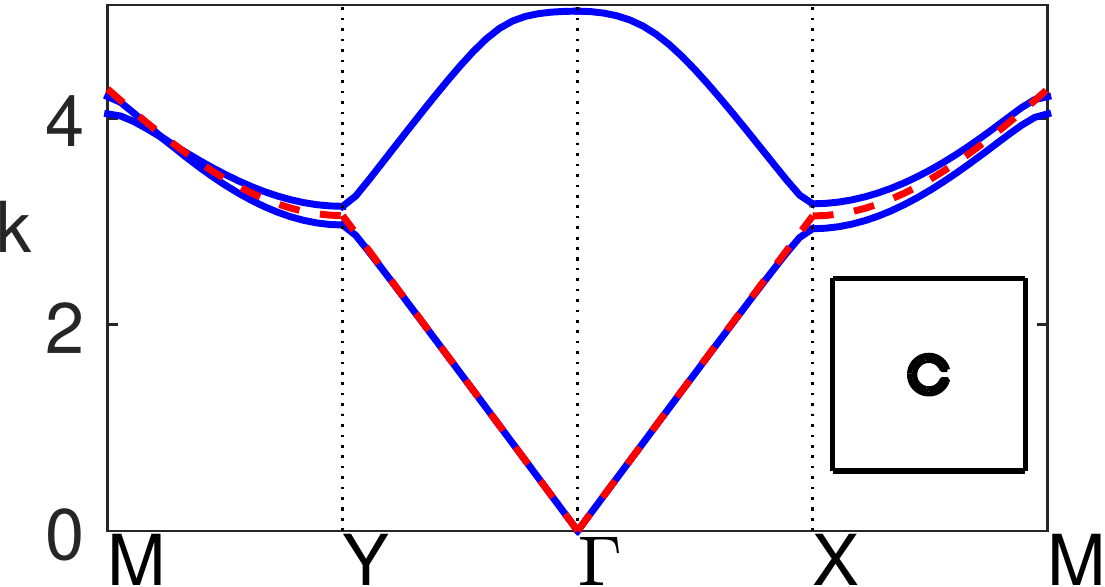}
\label{fig:thickmultipole12subfig1}}
\subfloat[Subfigure 1 list of figures text][]{
\includegraphics[width=0.475\textwidth]{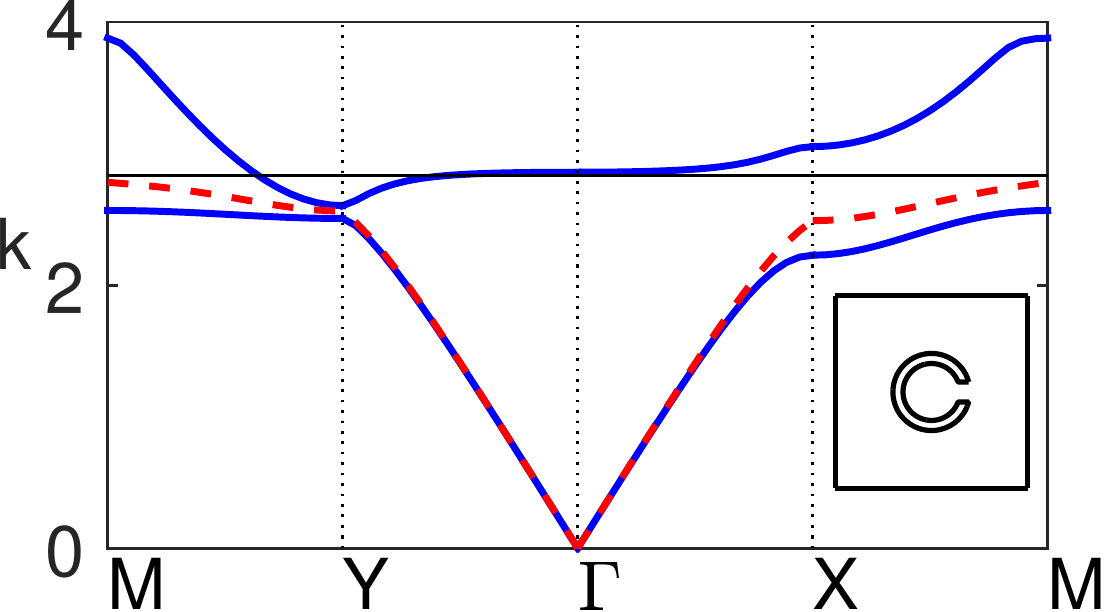}
\label{fig:thickmultipole12subfig2}}\\
\subfloat[Subfigure 2 list of figures text][]{
\includegraphics[width=0.475\textwidth]{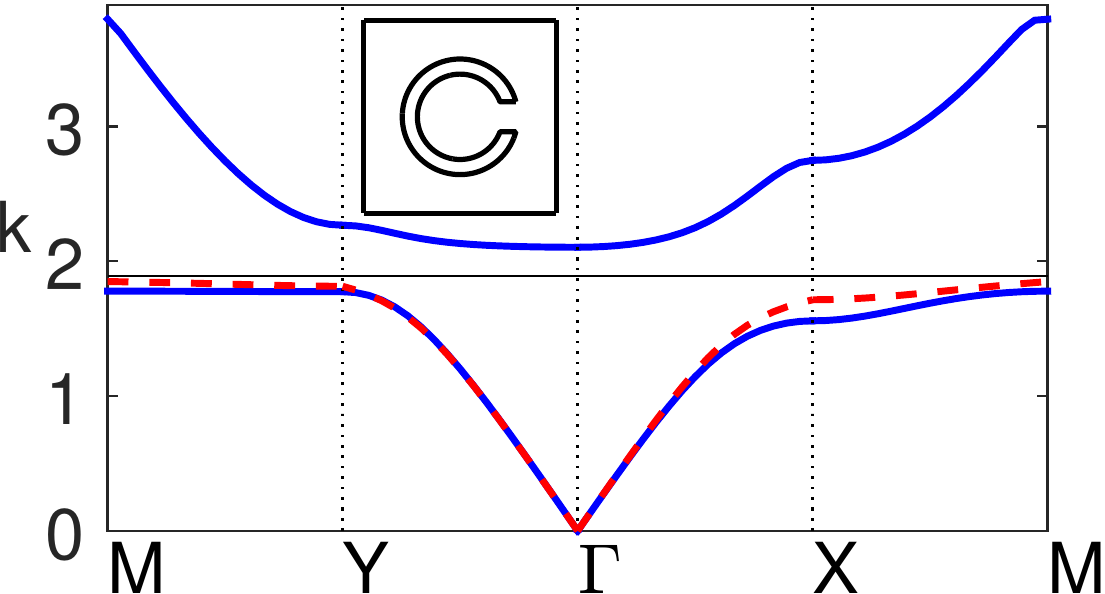}
\label{fig:thickmultipole12subfig3}}
\subfloat[Subfigure 3 list of figures text][]{
\includegraphics[width=0.475\textwidth]{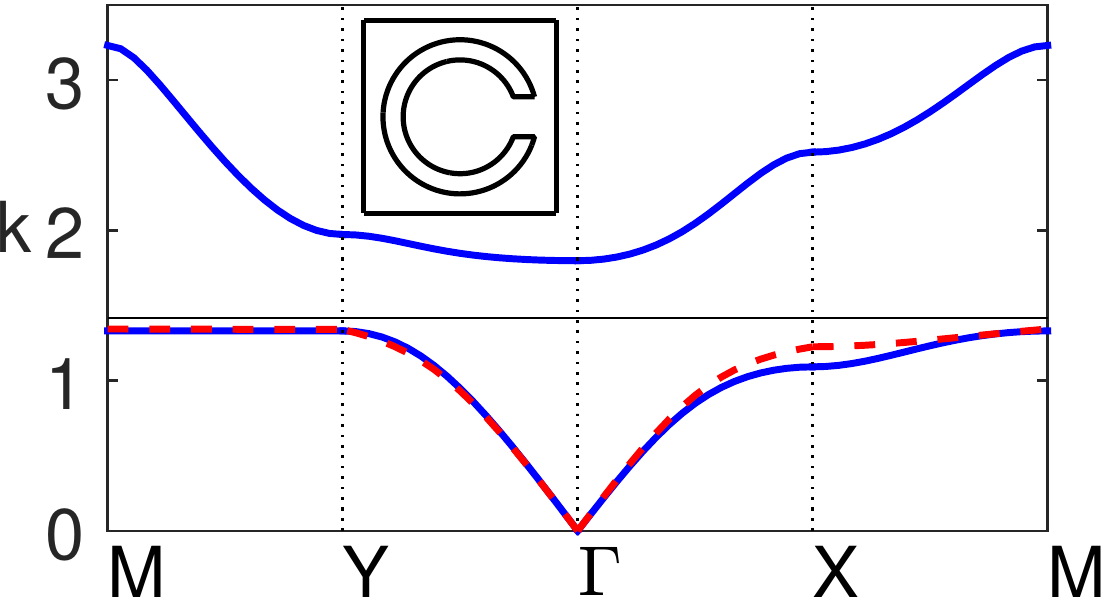}
\label{fig:thickmultipole12subfig4}}
\caption{Band diagrams for a two-dimensional square array of thick-walled Helmholtz resonators  as  the  outer radius $b$ (equiv.~filling fraction) is increased for fixed neck length ratio   $h=0.5$:  \protect\subref{fig:thickmultipole12subfig1}  $b=0.1$, \protect\subref{fig:thickmultipole12subfig2} $b=0.2$,   \protect\subref{fig:thickmultipole12subfig3} $b=0.3$, and  \protect\subref{fig:thickmultipole12subfig4} $b=0.4$, with fundamental unit cells inset. Blue lines denote results from finite-element methods, red dashed lines denote results from the first-order correction \eqref{eq:dispeqfirstorderthick}, and black lines denote estimates for the edge of the band gap \eqref{eq:maxeval1stbandthick}.     In the above figures we use $\bar{d} = 1$, $\theta_0 = 0$,    and $\theta_\mathrm{ap} = \pi/12$.}
\label{fig:thickness12ffracvariation}
\end{figure}

Finally, we consider the relationship between resonator wall thickness and filling fraction in Figure \ref{fig:thickness12ffracvariation} where we impose a channel width aspect ratio of $h=0.5$ and vary the outer radius $b$ for the aperture width $\theta_\mathrm{ap} = \pi/12$. We also superpose the first-order approximation for the first band  \eqref{eq:hepsfullthick}, and the first band maximum \eqref{eq:maxeval1stbandthick}. As in the thin-walled case, we observe a widening of the first band surface with increasing filling fraction (outer radius); this result may prove useful in countering the effect of thickness in the event that a wide band gap is sought. That is, although the presence of thickness may close the band gap it may be possible to compensate against this by tuning the outer radius. We find that the first band description generally works  well, with the exception of configurations where the first  and second band are almost degenerate at the $Y$ point, and at this wider half-angle $\theta_\mathrm{ap} = \pi/12$ we observe a slight loss of accuracy in the band curvature near the   saddle point at $X$.

  \begin{figure}[t]
\centering
\subfloat[Subfigure 6 list of figures text][]{
\includegraphics[width=0.475\textwidth]{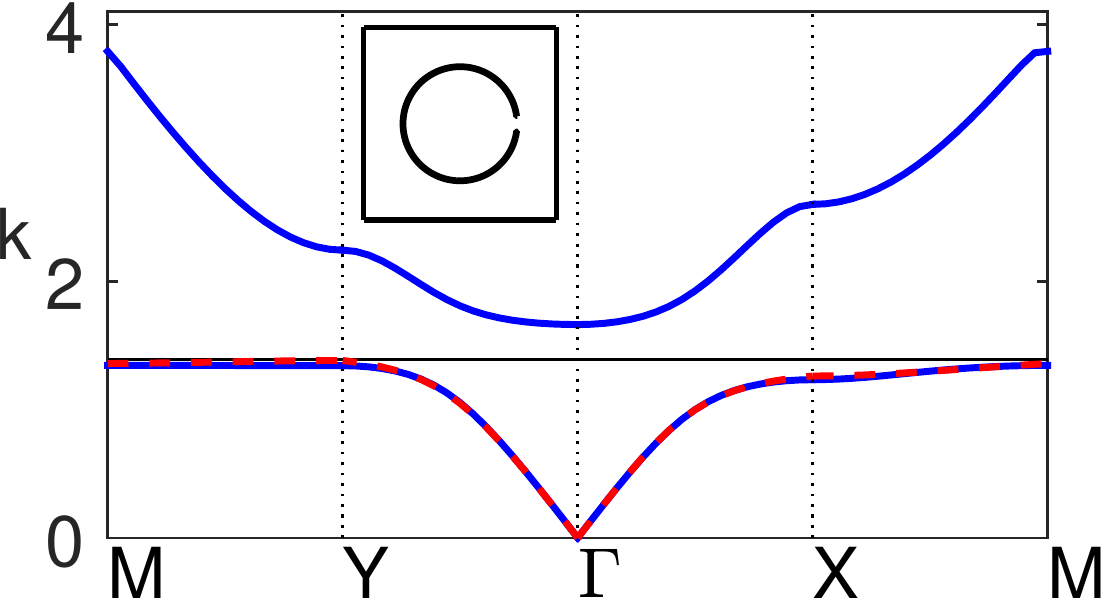}
\label{fig:thickmultipole24subfig1}}
\subfloat[Subfigure 1 list of figures text][]{
\includegraphics[width=0.475\textwidth]{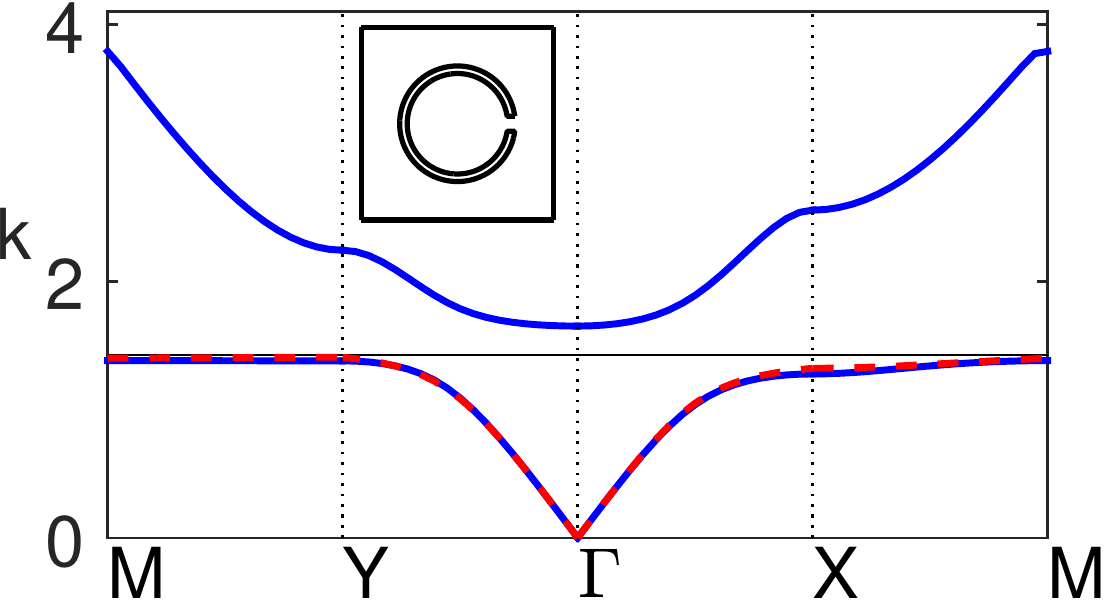}
\label{fig:thickmultipole24subfig2}}\\
\subfloat[Subfigure 2 list of figures text][]{
\includegraphics[width=0.475\textwidth]{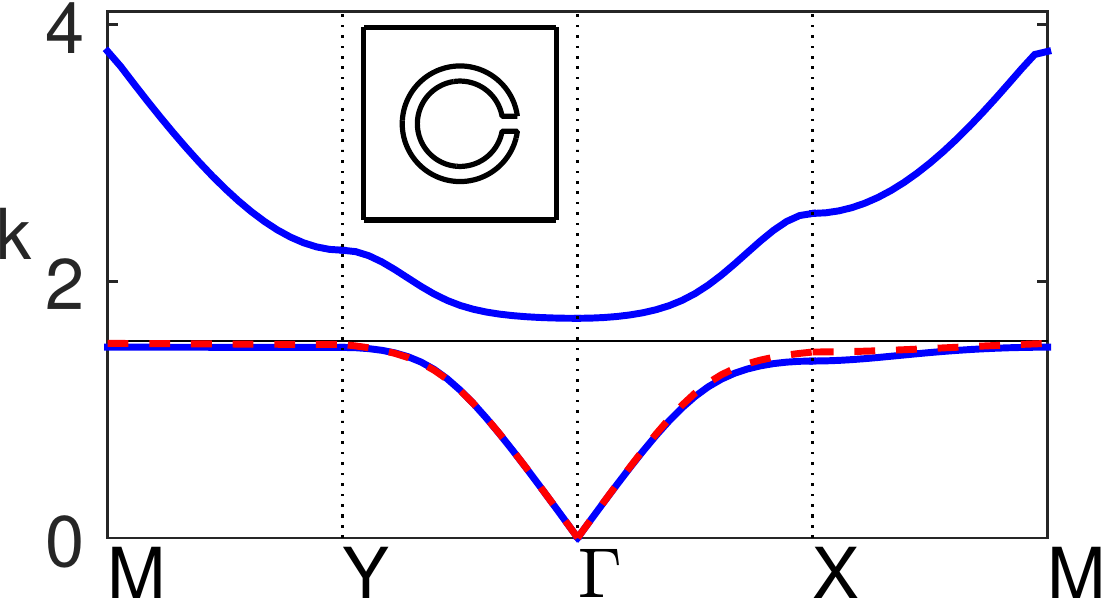}
\label{fig:thickmultipole24subfig3}}
\subfloat[Subfigure 3 list of figures text][]{
\includegraphics[width=0.475\textwidth]{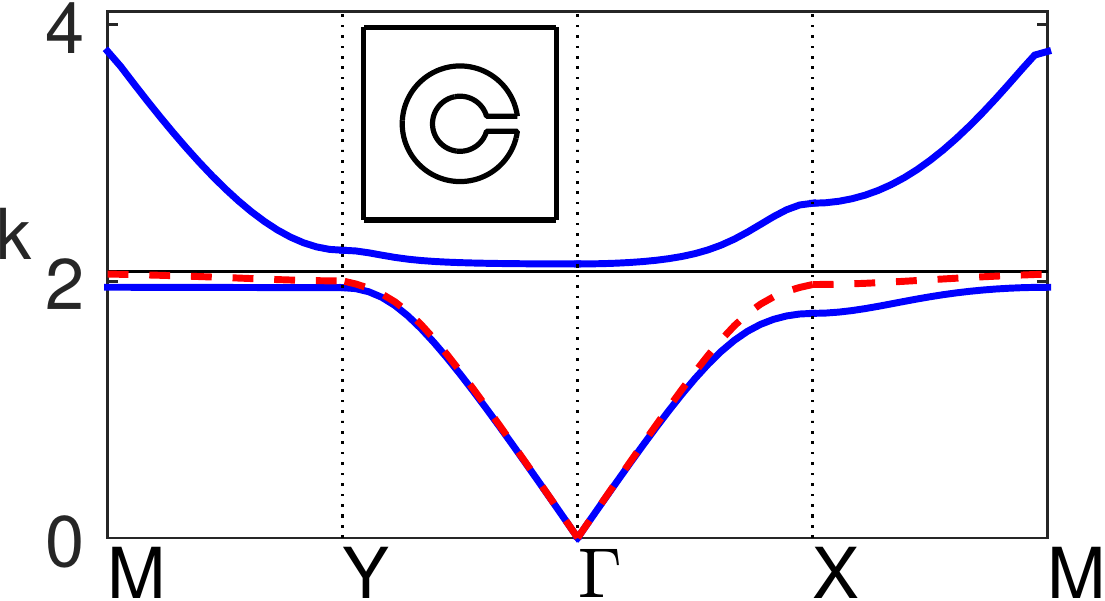}
\label{fig:thickmultipole24subfig4}}
\caption{Band diagrams for a two-dimensional square array of thick-walled Helmholtz resonators  as  the wall thickness $h$   is increased:  \protect\subref{fig:thickmultipole24subfig1}  $h=0.1$, \protect\subref{fig:thickmultipole24subfig2} $h=0.5$, \protect\subref{fig:thickmultipole24subfig3} $h=1$, and \protect\subref{fig:thickmultipole24subfig4} $h=2$, with fundamental unit cells inset. Blue lines denote results from finite-element methods, red dashed lines denote results from the first-order correction \eqref{eq:dispeqfirstorderthick}, and black lines denote estimates for the edge of the band gap \eqref{eq:maxeval1stbandthick}.  In the above figures we use $\bar{d} = 1$, $\theta_0 = 0$,   $\bar{b} = 0.3$, and $\theta_\mathrm{ap} = \pi/24$.}
\label{fig:thickness24hvariation}
\end{figure}

  \begin{figure}[t]
\centering
\subfloat[Subfigure 6 list of figures text][]{
\includegraphics[width=0.475\textwidth]{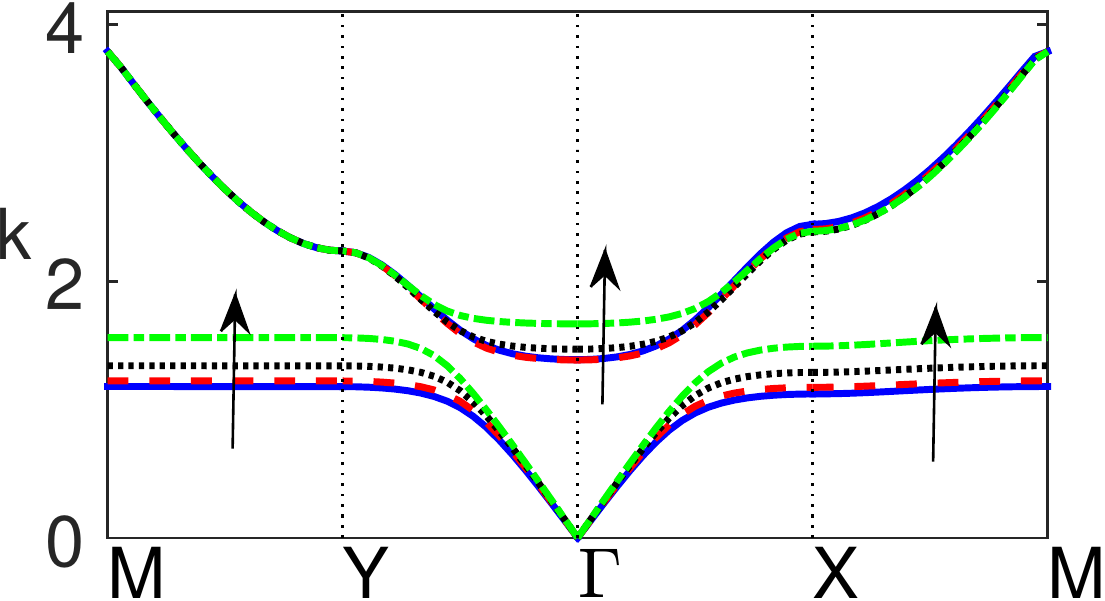}
\label{fig:comp1}}
\subfloat[Subfigure 1 list of figures text][]{
\includegraphics[width=0.475\textwidth]{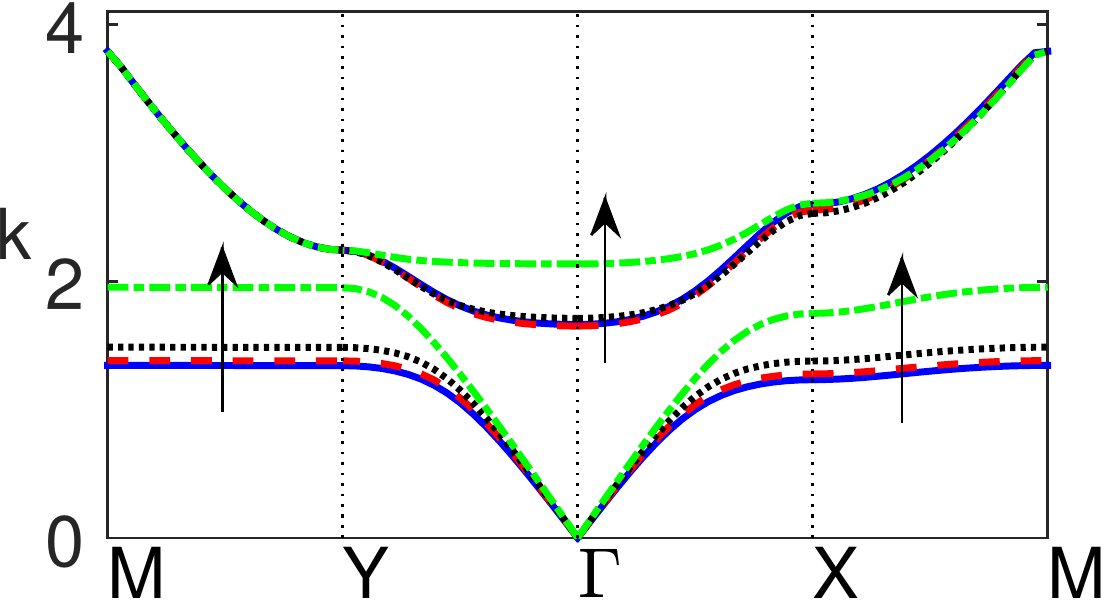}
\label{fig:comp2}} 
\caption{Superposition of the band diagrams in   \protect\subref{fig:comp1} Figure  \ref{fig:thickness48hvariation} and   \protect\subref{fig:comp2} Figure  \ref{fig:thickness24hvariation}, where arrows denote directions of increasing wall thickness $h$. }
\label{fig:compallfigs}
\end{figure}

 \section{Concluding remarks} 
 \label{chap:concl}
 We have  constructed a multipole formulation for calculating the band structure of a medium comprising a two-dimensional square array of  thick- and thin-walled Helmholtz resonators embedded in a uniform background. The eigenvalue problem was derived  using   both multipole methods and the method of matched asymptotic expansions,  from which we were  able to  extract a  dispersion equation approximation analytically that implicitly defines the first band surface over the entire Brillouin zone. For thin-walled resonators we find that both the multipole formulation and the first-band surface description perform surprisingly well over a wide selection of aperture widths and filling fractions, compared to results from finite-element methods.  Likewise, for thick-walled resonators we find similarly strong performance across a selection of aperture widths and resonator neck thicknesses. A key feature of these Helmholtz resonator arrays is the emergence of a low-frequency band gap, where plane wave propagation through the array is not possible in the bulk material. We find that thin-walled resonators generally possess the widest gaps, and therefore for soundproofing applications   recommend making the resonator walls as thin as practicably possible. The formulation we present also makes it possible to conveniently determine configurations that return a desired phase and/or group velocity at long wavelengths, should this   be required.  We anticipate that our multipole--matched asymptotic formulation will prove useful beyond the field of acoustics,  such as in electromagnetism,  after a simple replacement  of constants (i.e., $B\mapsto \varepsilon_\rmr^{-1}$ and $\rho \mapsto \mu_\rmr$ \cite{movchan2002asymptotic}). The multipole-matched asymptotic expansion treatment outlined here provides closed-form expressions for the dispersion relation over a wide frequency range, which is  particularly valuable, since it may   be used to rapidly search over large parameter spaces   for optimal configurations.  Finally, we emphasise that the first band descriptions obtained   extend   outside the classical  long wavelength regime, and are therefore useful for describing how plane waves propagate through the array over very large frequency ranges.

\appendix

 \section{Convergent lattice sum definition}
 \label{chap:SlY}
The  lattice sums  $S_\ell^\rmY$ are most often    defined via   the conditionally convergent form \cite{movchan2002asymptotic}
 \begin{equation}
\label{eq:Slylatt}
S_\ell^\rmY(\bfk_\rmB) = \sum_{m,n}^{} {}^\prime \rmY_\ell ( {R}_{mn}) \rme^{\rmi \ell \phi_{mn}} \rme^{\rmi \bfk_\rmB \cdot \mathbf{R}_{mn}},
\end{equation}
where   $\mathbf{R}_{mn} = R_{mn} \, \mathrm{exp}(\rmi \phi_{mn})$ is the (dimensionless) lattice generator in polar coordinates (i.e., $\bfR_{mn} =( dm , dn)$ for a square lattice of period $d$ and where $m,n \in \mathbb{Z}$),   $\bfk_\rmB = (k_{\rmB x}, k_{\rmB y})$ is the dimensionless Bloch vector, and  prime notation denotes summation over all points in the array excluding $m=n=0$. We remark that there are many ways  in which this conditionally convergent sum may be regularised to obtain an absolutely convergent form \cite{linton2010lattice}; we present the well-known expression for a square lattice   as \cite[Eq. (3.104)]{movchan2002asymptotic} 
 \begin{multline}
 \label{eq:convergentSmY}
S_m^\rmY(\bfk_\rmB)  = \frac{1}{\rmJ_{m+r}(  \xi)} \left(- \left[ \rmY_r( \xi) + \frac{1}{\pi} \sum_{n=1}^{r} \frac{(r-n)!}{(n-1)!} \left( \frac{2}{  \xi}\right)^{r-2n+2}\right]\delta_{m,0} \right.\\
\left.- \frac{4  \I^m}{d^2}\sum_{p,q} \left( \frac{1}{Q_{pq}}\right)^r \frac{\rmJ_{m+r}(Q_{pq} \,  \xi)}{Q_{pq}^2 - 1} \E^{\I m \theta_{pq}} \right),
\end{multline}
where $r$ and $\xi$ are   regularisation parameters, with $r$ denoting  a small non-negative integer (e.g., $r=3$) and $\xi$   a small positive number which formally limits to zero (e.g., $\xi = d/100$). We also define the reciprocal lattice generator   for a square lattice $\bfK_{pq} = \left(2\pi p/d , 2 \pi q /d \right)$   where $p,q \in \mathbb{Z}$   and the translated reciprocal lattice generator $\bfQ_{pq} = \bfK_{pq} + \bfk_\rmB = Q_{pq} \exp(\rmi \theta_{pq})$.

\section*{Acknowledgements}
   I.D.A. acknowledges  support from a Royal Society Industry Fellowship. This work was also supported by EPSRC grant no EP/R014604/1 whilst I.D.A. held the position of Director of the Isaac Newton Institute Cambridge.

\bibliographystyle{RS.bst}

\end{document}